\documentclass[sigconf]{acmart}

\def\BibTeX{{\rm B\kern-.05em{\sc i\kern-.025em b}\kern-.08emT\kern-.1667em\lower.7ex\hbox{E}\kern-.125emX}}

\copyrightyear{2019}
\acmYear{2019}
\setcopyright{acmlicensed}
\acmConference[ASIACCS '19]{ASIACCS '19: ACM Symposium on AISACCS}{June 03--05, 2019}{ASIACCS, AKL}
\acmBooktitle{ASIACCS '19: ACM Symposium on AISACCS, June 03--05, 2019, AKL}
\acmPrice{15.00}
\acmDOI{10.1145/1122445.1122456}
\acmISBN{978-1-4503-9999-9/18/06}

\usepackage{booktabs} 
\usepackage{listings}
\usepackage{xcolor}
\usepackage{subfigure}
\usepackage{graphics}

\usepackage{color}
\definecolor{light-gray}{gray}{0.80}

\begin{document}

\title{A Feature-Oriented Corpus for Understanding, Evaluating and Improving Fuzz Testing}
\author{Xiaogang Zhu, Xiaotao Feng, Tengyun Jiao}
\email{xiaogangzhu@swin.edu.au}
\affiliation{%
  \institution{Swinburne University of Technology}
  \city{Melbourne}
  \state{VIC}
  \postcode{3122}
}

\author{Sheng Wen, Yang Xiang}
\email{swen,yxiang@swin.edu.au}
\affiliation{%
  \institution{Swinburne University of Technology}
  \city{Melbourne}
  \state{VIC}
  \postcode{3122}
}

\author{Seyit Camtepe}
\email{Seyit.Camtepe@data61.csiro.au}
\affiliation{%
 \institution{DATA61 | CSIRO}
  \city{Sydney}
  \state{NSW}
  \postcode{2122}
}

\author{Jingling Xue}
\email{jingling@cse.unsw.edu.au}
\affiliation{%
  \institution{The University of New South Wales}
  \city{Sydney}
  \state{NSW}
  \postcode{2052}
}

\renewcommand{\shortauthors}{XG Zhu, et al.}

\begin{abstract}
Fuzzing is a promising technique for detecting security vulnerabilities.
Newly developed fuzzers are typically evaluated in terms of the number of bugs found on vulnerable programs/binaries. However, existing corpora usually do not capture the features that prevent fuzzers from finding bugs, leading to ambiguous conclusions on the pros and cons of the fuzzers evaluated. A typical example is that Driller \cite{Stephens2016Driller} detects more bugs than AFL \cite{aflweb}, but its evaluation cannot establish if the advancement of Driller stems from the concolic execution or not, since, for example, its ability in resolving a dataset's magic values is unclear.

In this paper, we propose to address the above problem by generating corpora based on search-hampering features. As a proof-of-concept, we have designed FEData, a prototype corpus that currently focuses on four search-hampering features to generate vulnerable programs for fuzz testing. Unlike existing corpora that can only answer ``how'', FEData can also further answer ``why'' by exposing (or understanding) the reasons for the identified weaknesses in a fuzzer.
The ``why'' information serves as the key to the improvement of fuzzers.

To show the utility of FEData, we evaluate two representative fuzzers, AFL \cite{aflweb} and AFLFast \cite{Boehme2016Coverage}, on FEData. Our results confirm that AFLFast is indeed faster than AFL in terms of path search, but at the expense of missing bugs that can be found by AFL in some programs with ``deep'' bugs and/or few dataflows. Our FEData programs have enabled us to identify the root cause, called \emph{cycle explosion}, behind (for the first time).
Based on this new finding, we have developed an improved version of AFLFast, called \texttt{AFLFast+}. Our evaluation demonstrates that AFLFast+ has overcome the cycle explosion problem, by retaining the efficiency of AFLFast in path search while also maintaining or even surpassing the bug-finding capability of AFL for the corpus evaluated.

\end{abstract}

%


%

\maketitle

\section{Introduction}
Fuzzing is an automatic software testing technique that typically provides random data as inputs to programs and then monitors them for exceptions such as crashes. Fuzzing can capture bugs\footnote{Vulnerability is different from bug. Bug brings a program into an unintended state. When a bug can be exploited by an attacker, the bug becomes a vulnerability \cite{Muench2018What}. In the following of this paper, we will use the word ``bug'' instead of ``vulnerability'' for brevity, but what we mean is ``vulnerability''.} because the exceptions are usually the indicators of bugs in the program context.
In 1990, B.P. Miller \textit{et al.} \cite{Miller1990empirical} developed the first fuzzing algorithm (fuzzer for short). Since then, fuzzing has become one of the major tools for detecting bugs.

The performance of a fuzzer is mainly determined by its capability of handling search-hampering features in the contexts of bugs. For example, if a bug is `protected' by magic values (\textit{e.g.}, $1$ in an \texttt{if} statement `\texttt{if (x==1)}'), a fuzzing algorithm has to resolve every specific value before it triggers the bug. This means that a fuzzer that resolves more magic values has a higher chance in examining more vulnerable programs and more bugs. In addition, bugs can be hidden in a large number of execution paths in a vulnerable program. Hence, a fuzzer has better performance if it verifies execution paths for bugs more efficiently. In these two situations, the number of magic values and execution paths are considered as the major factors that prevent fuzzing algorithms from exposing the underlying bugs. As search-hampering features are important to the fuzzing performance, most fuzzers \cite{Pham2016Model,Wang2017Skyfire,Corina2017Difuze,Li2017Steelix,Peng2018T,Stephens2016Driller,Boehme2016Coverage,Gan2018CollAFL,Rawat2017Vuzzer} are tailored to resolve one/several features to increase the coverage of code inspection.

Though many fuzzers were developed, their appropriate evaluation is still a big challenge due to the lack of supportive corpora. To validate a newly developed fuzzer, a corpus needs to contain the contexts of bugs such as specific search-hampering features for fuzzing.
Those contexts can help validate the advancement of a fuzzer resolving the specific challenges.
However, it is not feasible for fuzzing so far. In current stage, the usual way to evaluate fuzzers is to run them on real-world program corpora, and judge their performance by counting the number of unique crashes after a period of running time (normally 24 hours) \cite{Chen2018Angora,Boehme2016Coverage,Corina2017Difuze,Haller2013Dowsing,Schumilo2017kAFL,Wang2017Skyfire,Peng2018T}. As disclosed by G. Klees \textit{et al.} \cite{Klees2018Evaluating}, this often leads to ambiguous or even wrong conclusions on the fuzzers, since it is almost impossible to provide supportive contexts of bugs by inspecting a large number of programs in corpora. A simple example is that, Driller \cite{Stephens2016Driller} detects more bugs than AFL \cite{aflweb} by utilizing concolic execution to resolve magic values. But the evaluation results may not support the advancement of Driller when the corpora do not have enough bugs `protected' by magic values. In fact, even though we reckon the evaluation results are convincing, we still cannot pinpoint the reason that leads to fuzzers' advancements or weaknesses without contextual details of bugs in corpora. This hampers or even disables our attempts on improving existing fuzzers.

So far, researchers have already made some efforts to address the above evaluation challenge. For example, some researchers prefer to create corpora based on real-world programs, typically from student code \cite{spacco2005bug}, existing bug report databases \cite{lu2005bugbench}, or by creating a public bug registry \cite{foster2005call,FuzzerTestSuite2018Fuzze}. Despite these proposals provide corpora contextual details of bugs, they have remained static and relatively small for fuzzing evaluation. There are also some tries on creating independently defined public benchmark suites, \textit{e.g.} DaCapo \cite{blackburn2006dacapo} and SPEC \cite{specweb}. Among these attempts, even though they collect a large volume of real-world vulnerable programs, it is still painstaking to manually triage the crashes and filter the bugs in those programs. Before a strong community effort has been applied to these suites, it is believed that they are not sufficient to support fuzzing evaluation \cite{Klees2018Evaluating}.

An alternative way to create fuzzing corpora is to synthesize vulnerable programs. Typical examples include early corpora for buffer overflow detection \cite{wilander2003comparison, zitser2004testing}, LAVA and next version LAVA-M \cite{Dolan-Gavitt2016LAVA}, DARPA CGC corpus \cite{CGCCORPUS}, as well as corpus drawn from NIST SAMATE project \cite{nistJuliet}. The synthetic corpora ensure the existence of bugs by sacrificing their reflection on real-world ones. However, to the best of our knowledge, none of the above synthetic corpora have implemented contexts of bugs pertaining to search-hampering features. Therefore, synthetic corpora also cannot support fuzzing evaluation according to aforementioned analysis.

In this paper, we propose generating corpora based on search-hampering features to solve the above challenge. To this end, a framework is developed and used to synthesize evaluation corpora in an automatic manner. The vulnerable programs are made up by function-level structures from GitHub\footnote{https://github.com} code to maintain as much as possible the style of real-world programs. The framework ensures the generated programs are able to be compiled, and inserts the contexts of bugs to the programs when necessary. Holding contexts of bugs, the generated corpora can not only make the conclusions of a fuzzer's advancements solid, but also expose the clues for improving a fuzzer. To demonstrate the effectiveness of our idea, we develop a proof-of-concept corpus, namely \texttt{FEData}, based on four typical search-hampering features \textit{i.e.}, the dataflow to trigger bugs, the number of magic values, the number of execution paths, and the number of checksums.

As a case study, we run AFL and AFLFast \cite{Boehme2016Coverage} on programs from FEData to validate its utility in fuzzing evaluation. The results of our experiments confirm the conclusion made by AFLFast that it finds execution paths faster than AFL. The results also indicate that AFLFast detects fewer bugs than AFL in some specific programs from FEData. G. Klees \textit{et al.} \cite{Klees2018Evaluating} found similar phenomenon that AFLFast detected fewer bugs when they ran these two fuzzers more than 24 hours, but they did not explain the reason for it. With contexts of bugs on those programs, we find that AFLFast is prone to fall into the \texttt{cycle explosion} state. As AFLFast can no longer produce new inputs in this state, it will find fewer bugs than AFL. Accordingly, we improve AFLFast and develop \texttt{AFLFast+} by setting a lower bound to the number of inputs produced by AFLFast, which prevents fuzzing from dropping into the cycle explosion state.

We summarize the contributions as follows:
\begin{itemize}
\item We propose generating corpora for fuzzing evaluation based on search-hampering features. The generation runs in an automatic manner, and the generated corpora are not only used for understanding and evaluating fuzzing but also help improve the fuzzers based on the contexts of bugs.
\item We carry out a case study on AFLFast via a proof-of-concept corpus, called FEData. The corpus is generated based on four typical search-hampering features. The experimental results validates the utility of the feature-oriented corpus by showing its accurate conclusions in fuzzing evaluation.
\item We develop AFLFast+ according to AFLFast's weakness that we expose in the case study. The experiments show that AFLFast+ can detect more bugs than AFLFast when other factors stay the same.
\end{itemize}

\begin{figure}[thp!]
    \centering
    \includegraphics[width=1.0\columnwidth]{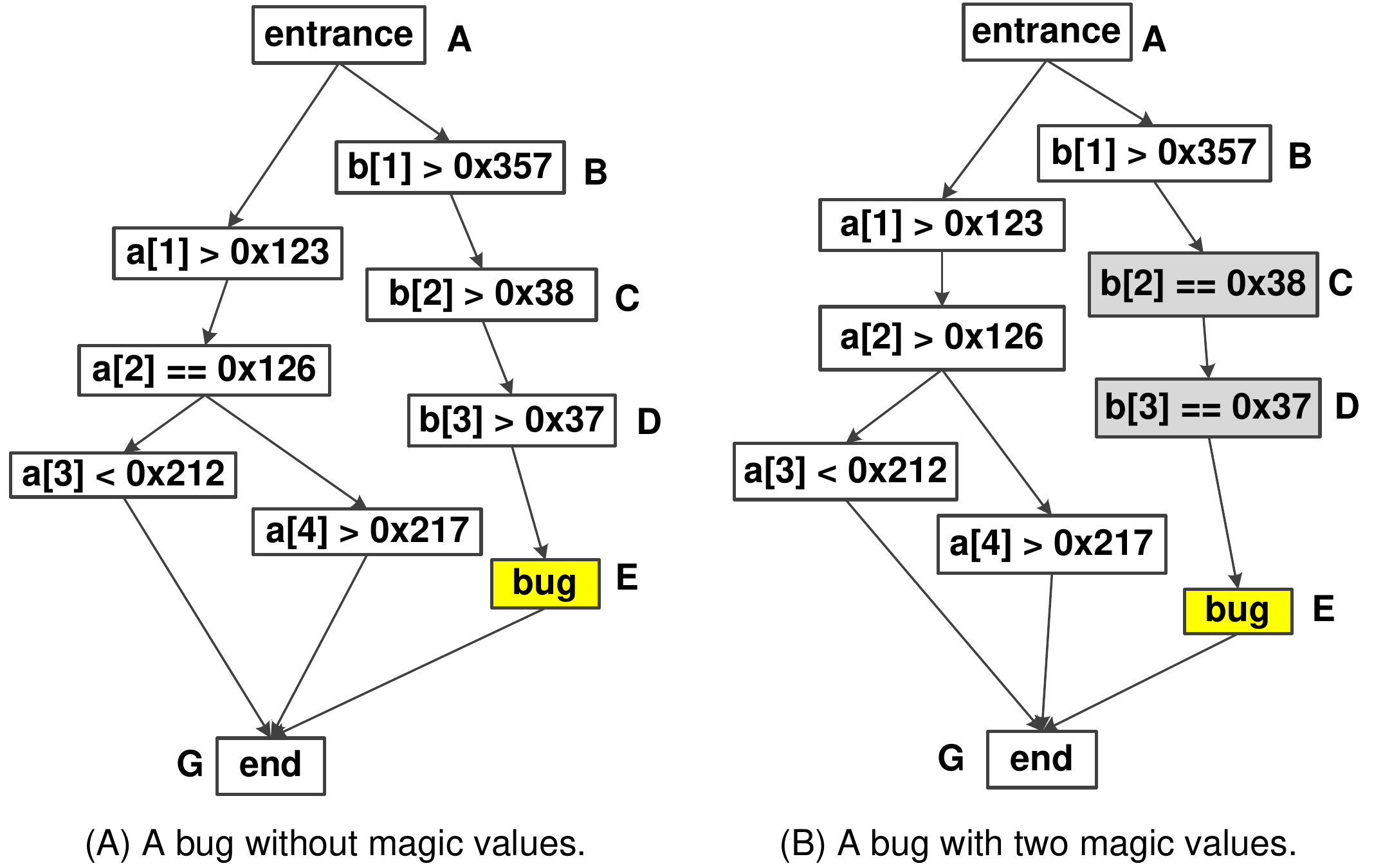}
    \caption{Two program examples that have different numbers of magic values to protect bugs. It may lead to a wrong conclusion if fuzzers such as Driller and AFL are evaluated on program (A). Driller outperforms AFL only on program (B) since bug is hidden behind two magic value challenges.}
    \label{fig-driller-exams}
\end{figure}

\begin{figure*}[t]
    \centering
    \includegraphics[width=0.92\textwidth]{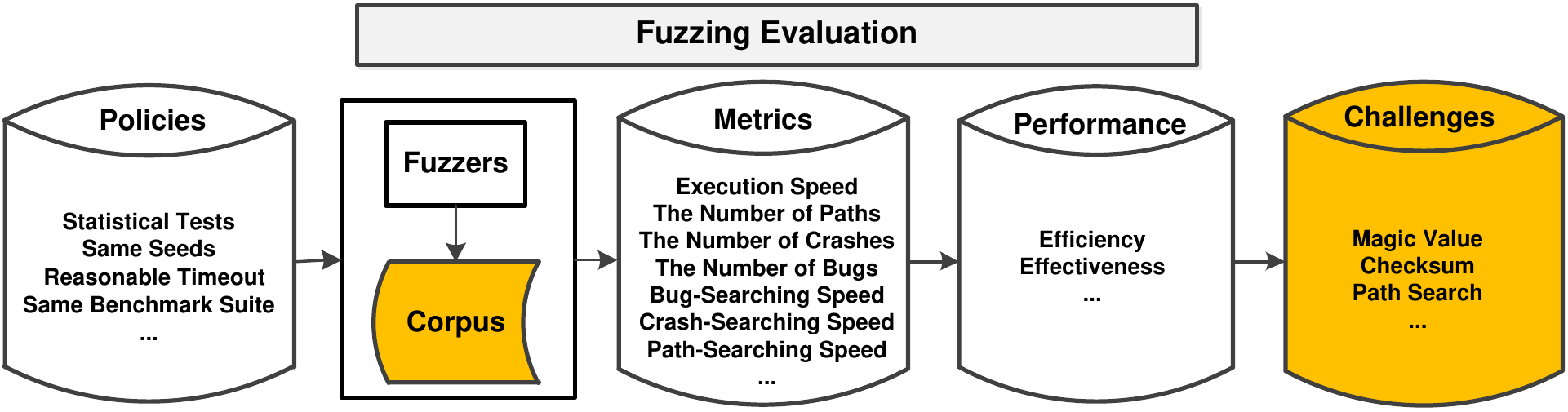}
    \caption{The procedure for evaluating fuzzers. Policies limit the way to run fuzz testing. The performance of a fuzzer can be evaluated based on the metrics, which is mainly determined by the capability of solving challenges.}
    \label{fig:fuzz-eval-proc}
\end{figure*}

\section{Overview of the Idea} \label{sec-feature-orient}
The idea of generating feature-oriented corpora will help solve the challenge of fuzzing evaluation. To make the idea clearer, we take two fuzzers to exemplify the necessity of holding contexts of bugs for fuzzing evaluation.

We take the evaluation of Driller \cite{Stephens2016Driller} against AFL \cite{aflweb} as an example. Driller claims to beat AFL because it can resolve magic values more efficiently. In fact, Driller is an updated version of AFL since it takes AFL as the core and adopts concolic execution \cite{sen2007concolic} to resolve magic values for larger coverage. However, the claim of Driller may not be supported in fuzzing evaluation sometimes. As shown in Fig.\ref{fig-driller-exams}(A), because there are no magic values on the execution path $A\rightarrow B\rightarrow C\rightarrow D \rightarrow E\rightarrow G$, the concolic execution in Driller makes no difference from AFL to trigger the bug. Even worse, the concolic execution in Driller takes more time to resolve the magic value $0x126$ in an execution path where there is no bug. Considering the randomness nature of fuzzing, it is highly possible for Driller to have a worse performance than AFL. However, Driller can outperform AFL on the programs similar to Fig.\ref{fig-driller-exams}(B), where there are lots of bugs hidden behind magic value challenges. Therefore, when state-of-art corpora do not provide contexts of bugs for evaluation, the results may lead to ambiguous or even wrong conclusions on fuzzers.



It is necessary to implement contexts of bugs such as search-hampering features in fuzzing evaluation corpora. As shown in Fig.\ref{fig:fuzz-eval-proc}, the corpus of vulnerable programs is one of the major components in fuzzing evaluation procedure. Fuzzers follow a policy such as how to generate seeds, and run on the evaluation corpus. The efficiency of fuzzing is mainly determined by the capability of the fuzzer solving contextual challenges in programs, such as magic values, checksums, path search, \textit{etc}. We can therefore measure the performance of fuzzers via a list of metrics such as the number of magic values resolved in the execution. However, to date, the usual way to judge the performance of a fuzzer is to count the number of bugs exposed within a fixed period of time. Because the number of exposed bugs may not be positively correlated to the challenges, it cannot always indicate the efficiency of a fuzzer. In addition, since the number of exposed bugs cannot be linked to the metrics without contexts of bugs in programs, the usual way becomes too superficial to diagnose the drawbacks of a fuzzer. This explains our technical motivation as well as the real need to develop fuzzing evaluation corpora that implement contexts of bugs such as search-hampering features in vulnerable programs.

We create a framework to automatically generate corpora for fuzzing evaluation. The framework uses real-world program structures from GitHub to maintain the synthetic coding style. The contexts of bugs is then inserted into the structure, followed by extra code that ensures the generated programs are able to be compiled. Different from existing synthetic corpora, the design of our evaluation corpora is determined by the type(s) of search-hampering features that the fuzzer focuses on. In other words, the design is more `feature-oriented' for fuzzing evaluation.

\section{Search-hampering Features} \label{sec-Search-hampering}
Search-hampering feature is the major factor that affects the fuzzing performance and their evaluations according to the above analysis.
In fact, there are many search-hampering features.
In this section, we only select and focus on four typical ones: 1) the dataflow to trigger bugs, 2) the number of execution paths, 3) the number of magic values, and 4) the number of checksums.
The dataflow to trigger bugs is a basic and important feature for fuzzing evaluation.
When a fuzzer crashes a program, the dataflow to trigger a bug can help determine whether the crash is caused by the bug or not.
The other three features are important due to the challenges that fuzzing attempts to solve. As a program consists of three fundamental control structures, which are sequence structure, decision structure, and loop structure, fuzzers have to find solutions for the following challenges, 1) checksums, 2) magic values, and 3) execution path search, in order to reach larger code coverage.
To date, many fuzzers \cite{Chen2018Angora,Gan2018CollAFL,Boehme2016Coverage,Stephens2016Driller,Chen2018Hawkeye,Li2017Steelix,Peng2018T,Rawat2017Vuzzer} have been proposed to improve the performance of solving these three challenges.

\begin{figure}[thp!]
\includegraphics[height=0.6\columnwidth]{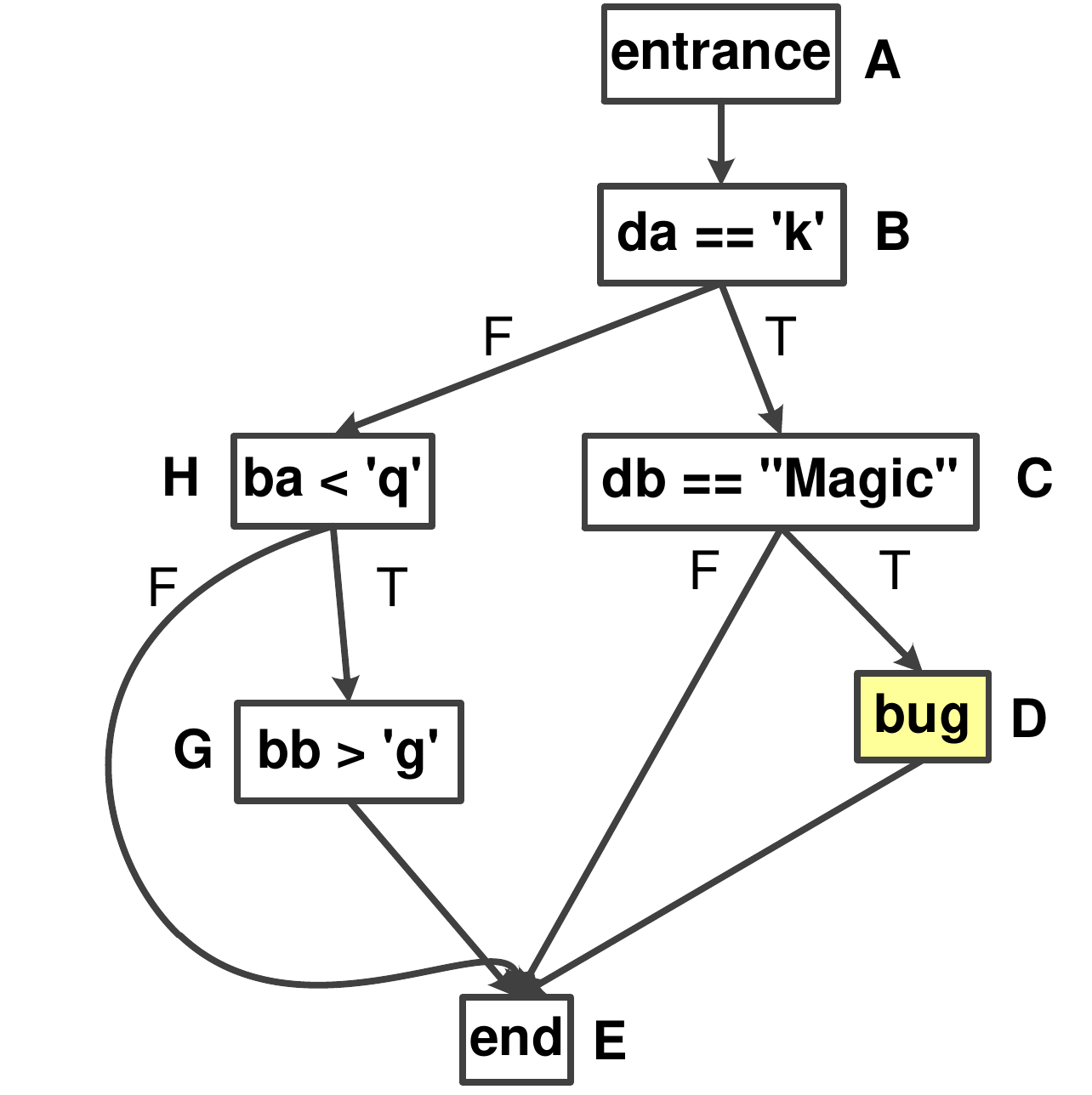} 
\caption{An example of the bug path and noise paths.
The bug path is $A\rightarrow B\rightarrow C\rightarrow D$ which can trigger the bug.
The noise paths cannot reach the bug, such as $A\rightarrow B\rightarrow H\rightarrow G\rightarrow E$.
`T' means True and `F' means False.}\label{m_spath_noise}
\end{figure}

\subsection{The Dataflow to Trigger Bugs}
A long dataflow will cost fuzzing more time than a short one to trigger bugs, because the fuzzing has to generate all the inputs in the dataflow.
Therefore, the dataflow to trigger bugs can affect the efficiency of fuzzing.
Moreover, a corpus that provides dataflows to trigger bugs will help count the bugs accurately.
As a crash is not equivalent to a bug, it may lead to a wrong conclusion about the efficiency of a fuzzer when the number of unique crashes are regarded as the number of bugs.
For example, suppose that two execution paths can reach the same bug, \textit{i.e.}, path $A\rightarrow B\rightarrow C\rightarrow D\rightarrow G$ and path $A\rightarrow E\rightarrow H\rightarrow G$.
Two different crashing inputs are generated by fuzzers but only one bug exists.

To implement dataflow that triggers bugs, a simple way is to insert a bug into one program and the execution path to trigger the bug is unique.
Therefore, we do not have to consider the side effects from multiple dataflows.
We call this unique execution path \texttt{Bug Path}.
All the execution paths that cannot reach bugs are called \texttt{Noise Paths}.
Fig.\ref{m_spath_noise} shows these two kinds of execution paths.
One of them is the bug path that triggers the bug, \textit{i.e.}, $A\rightarrow B\rightarrow C\rightarrow D$.
The other paths are the noise paths which cannot reach the bug block, such as $A\rightarrow B\rightarrow H\rightarrow G\rightarrow E$.
When conditions in block \textit{B} and \textit{C} are recorded in the corpus, they can be utilized to verify whether a crash is caused by the inserted bug.
If a dataflow satisfies the recorded conditions, the inserted bug is found.

\subsection{The Number of Execution Paths}\label{sbsec-num-paths}

\begin{figure}[thp!]
    \centering
    \includegraphics[width=0.95\columnwidth]{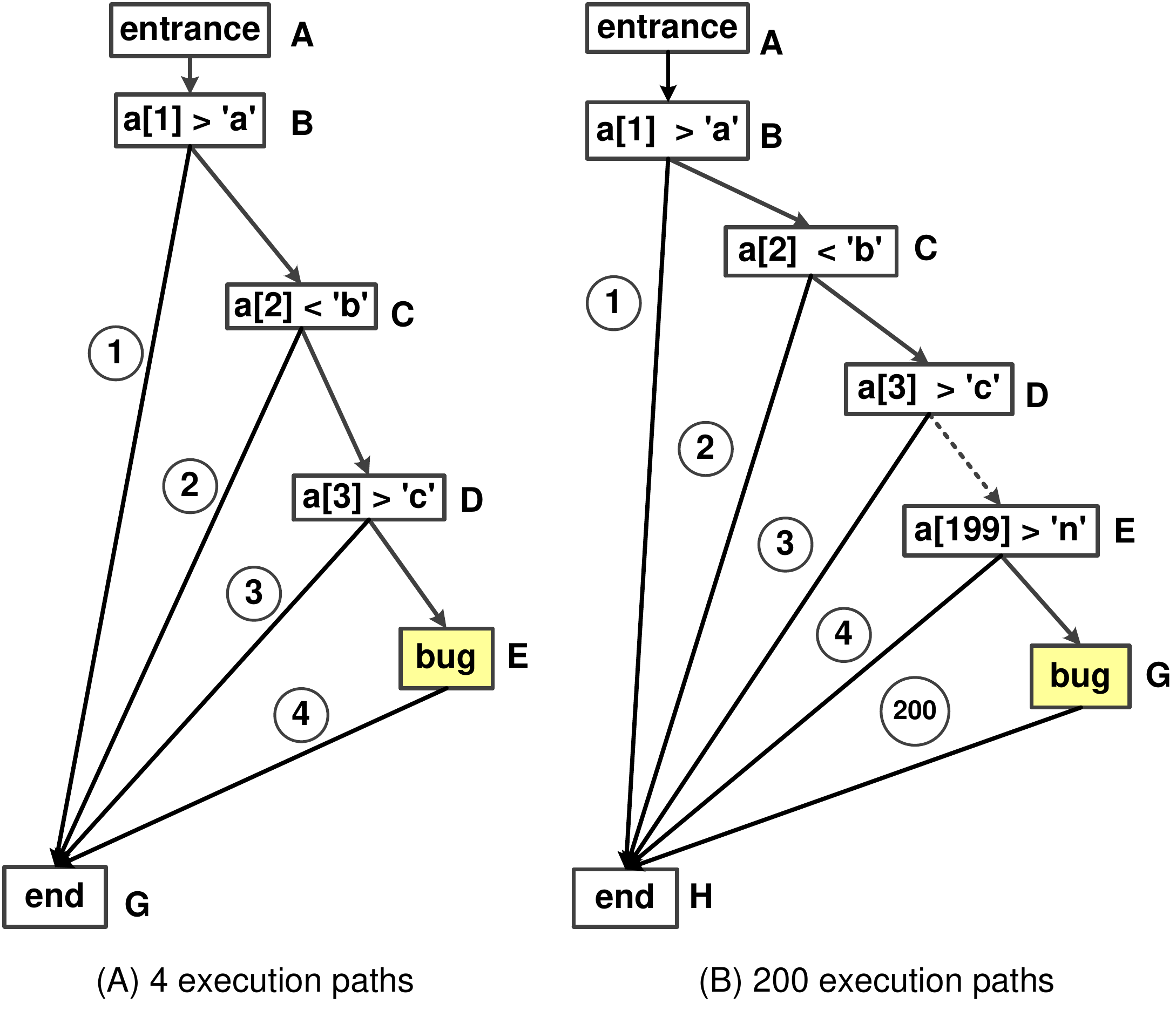}
    \caption{Different numbers of execution paths.
    It will cost fuzzing more time to locate the bug in (B) than in (A).}
    \label{fig-dif-paths}
\end{figure}

The number of execution paths affects the efficiency of fuzzing.
Fuzzing spends much time to locate the bug hidden in a large number of execution paths.
Fig.\ref{fig-dif-paths} shows different numbers of execution paths, where Fig.\ref{fig-dif-paths}(A) has four execution paths, and \ref{fig-dif-paths}(B) has 200 execution paths.
Fuzzing may spend a few seconds to find the bug hidden in \ref{fig-dif-paths}(A),
but it will cost fuzzing much more time to locate the bug hidden in \ref{fig-dif-paths}(B).

The execution paths in real-world programs are complicated, making it painstaking to count the number of executing paths.
In order to set the number of execution paths before a program is synthesized, one way is to restrict the control statements.
For example, if the following four conditions are satisfied, the number of execution paths will satisfy $p = c + 1$, where $p$ denotes the number of execution paths and $c$ denotes the number of input-related (\textit{i.e.}, variables getting their values from inputs) control statement conditions in the bug path.
First, the generated programs only have two control statements, \textit{if} and \textit{while}.
It is reasonable to have statements \textit{if} and \textit{while} because the combination of these two statements can substitute any combination of other statements.
Then, the conditions in \textit{while} statement do not contain any input-related variables and the loops will only repeat for fixed times.
This ensures that the generated programs will not include infinite loops.
Moreover, the inputs are only utilized by \textit{if} statement and each input is only used once.
Once this condition is satisfied, the bug path is feasible.
Finally, only the \textit{if} statement is utilized in the bug path.
Therefore, we can count the number of execution paths accurately.

For example, Listing \ref{gt_bug} presents the code implementation for Fig.\ref{fig-dif-paths}(A).
In this example, the dataflow to trigger the bug is (\textit{a[1] > "a", a[2] < "b", a[3] > "c"}).
In the bug path, the three \textit{if} statements have three independent inputs. In Fig.\ref{fig-dif-paths}(A), all the inputs are utilized in the bug path, and the number of input-related conditions is three while the number of execution paths is four.

\begin{lstlisting}[caption={Code for Fig.\ref{fig-dif-paths}(A). Each of the three values that are stored in the array $a$ is only used once in FEData.}, label=gt_bug, float=thp!]
int main(int argc, char* argv[]){
    char a[3];
    int n = 0;
    fgets(a,3,stdin);
    while (n < 2) n++;
    if (strcmp(a[1], 'a') > 0){
        if (strcmp(a[2], 'b') < 0){
            if (strcmp(a[3], 'c') > 0){
                bug();
            }
        }
    }
    return 0;
}
\end{lstlisting}

\subsection{The Number of Magic Values}
A magic value is a constant in the context of programs.
In fact, the nature of fuzzing is to generate inputs and find bugs automatically.
To detect bugs, fuzzers have to pass through checks and find the bug path.
Therefore, it will cost fuzzing more time to find bugs if the bugs are protected by magic values. We recall the example in Fig.\ref{fig-driller-exams}(B) to explain the magic values. In this example, the constants $0x38$ and $0x37$ are magic values as they are sitting in one side of the `==' conditions in the \textit{if} statements.
Moreover, because fuzzers regard every path as a potential bug path, they will attempt to resolve magic values in the noise paths as well.
Therefore, magic values are important to the efficiency of fuzzing, but it is not always true for a fuzzer to expose more bugs when it resolves magic values more efficiently.
As an example, Listing \ref{nested_con1} shows two magic values, wherein the magic value `MAGIC' does not protect the bug while the magic value `BYTE' does.
\begin{lstlisting}[caption={Magic values. Magic value `BYTE' protects the bug while `MAGIC' does not.}, label=nested_con1, escapechar=@, float=thp!]
int main(int argc, char ** argv){
  char inp1[6], inp2[6];
  fgets(inp1, 6, stdin);
  fgets(inp2, 6, stdin);
  @\colorbox{light-gray}{if(!strcmp(inp1, "MAGIC"))}@//magic value
  {
    printf("Not a bug.");
  }
  @\colorbox{light-gray}{if(!strcmp(inp2,"BYTE"))}@//magic value
  {
    bug();
  }
  return 0;
}
\end{lstlisting}

\subsection{The Number of Checksums}

A checksum is designed to detect errors in a block of digital data.
The effect of checksum is similar to the magic value but resolving a checksum is more complex because checksum needs complicated calculations.
A checksum function is a function that helps process the procedure of calculations.
This function can be utilized in a control statement condition, such as \textit{if}(\textit{func\_checksum}(\textit{a})).
In this scenario, a checksum can be regarded as a more complicated magic value.
For example, Listing \ref{checksum_lst} shows a simple checksum function.
In this function, two \textit{if} statements are introduced to check if specific conditions have been satisfied: 1) `$length(a)!=7$' and 2) `$sum(a)\%8==3$'.
We can see from this example that, to resolve a checksum is more complicated than to resolve a magic value.

\begin{lstlisting}[caption={A simple example of checksum function.
Two steps are taken to ensure that the variable $a$ satisfies the specific situations.}, label=checksum_lst, float=thp!]
bool func_checksum(int a[7]){
    if (length(a)!=7) return False;
    if (sum(a)%8==3){
        return True;
    }
    else{
        return False;
    }
}
\end{lstlisting}

\section{Proof-of-Concept: FEData} \label{sec-design}

We design a prototype corpus to show the effectiveness of our idea.
We narrow our focus on C programming language, which means all the programs in FEData are generated in C.
We will rewrite almost all the source code downloaded from the real-world programs and keep only the structure of them.

\begin{figure}[!htb]
\includegraphics[width=1.0\columnwidth]{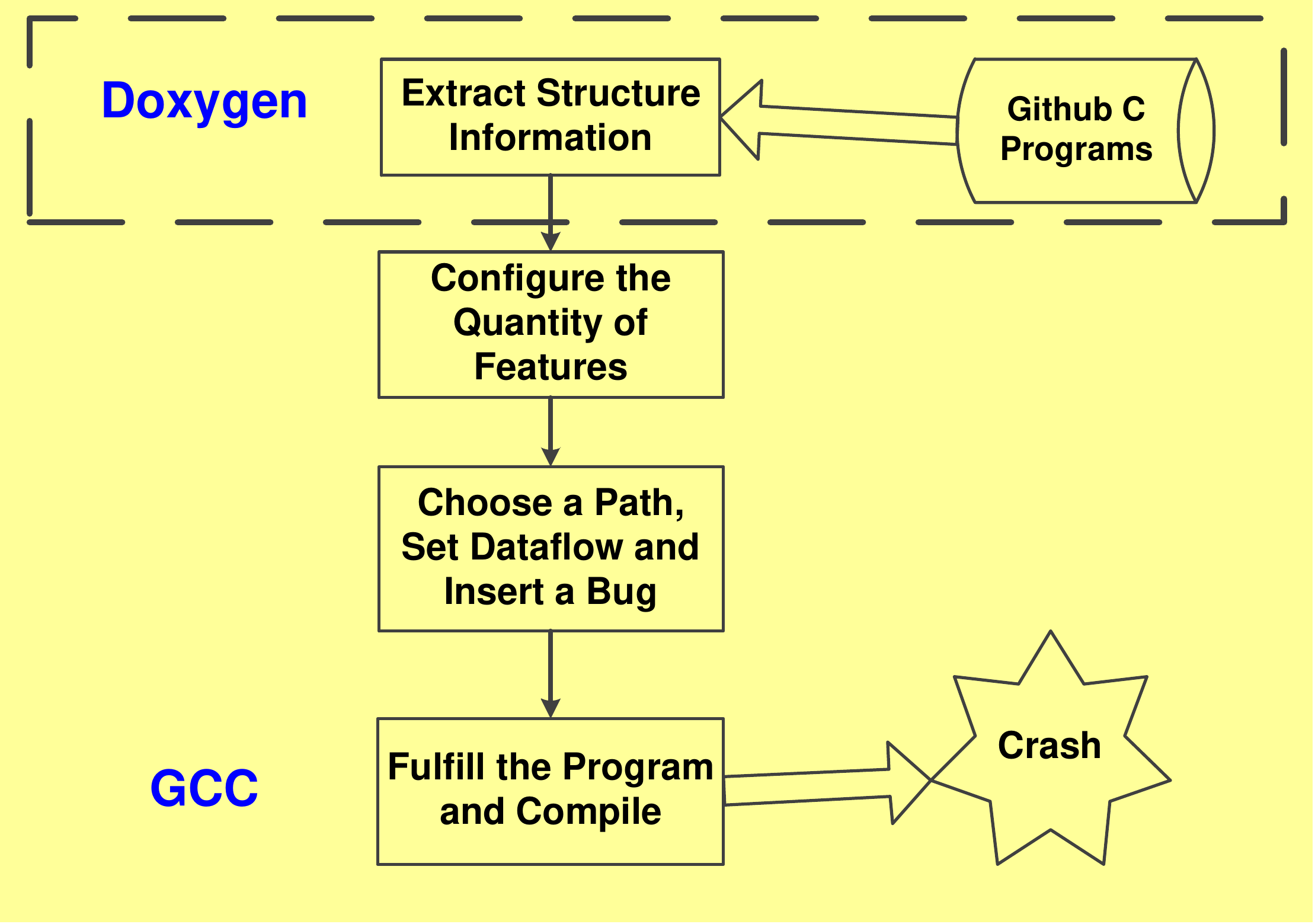} 
\caption{The design of FEData. The original programs are cloned from Github, and then they are transformed into FCGs.
Functions are simplified to insert our code.
Then, we choose a path in FCG to insert the bug.
Doxygen is used to extract information about the FCGs.}\label{prog_gene}
\end{figure}

It operates in four stages to generate FEData.
Fig.\ref{prog_gene} shows the design of FEData.
We first download many existing C programs and transform them into data that are suitable for generating our corpus.
Specifically, we extract function call graphs (FCGs) leveraging Doxygen\footnote{http://www.doxygen.nl/index.html} and Networkx\footnote{https://networkx.github.io}.
The functions are also simplified to keep only the structure we need.
Then, the context of a bug and the bug path are inserted into the FCGs.
Finally, we insert extra code to ensure that the generated programs are able to be compiled.

\subsection{Extract the Structure Information} \label{sec_extract_stru}

We download about 18,000 C programs from Github and use Doxygen and Networkx to extract FCGs from them.
To automatically generate a program, we randomly choose one of the 18,000 programs and add code.
Doxygen helps us rebuild the FCGs, but they cannot be used immediately.
We regard each FCG as a directed graph, in which a function is a vertex.
For the simplicity of the graph, we then use Networkx to break the cycles in the graph.
Networkx is also used to check if all the nodes in the graph are connected to the \textit{main} node (\textit{i.e.,} the \textit{main} function) and to remove the nodes that have no connection to the \textit{main} node.
This helps generate an executable program.

The functions are then simplified to insert search-hampering features.
Only semicolons, braces and keywords of control statements are used to generate new programs.
Specifically, we will add one code line for each semicolon in the simplified functions.
Moreover, the control statement keywords only include \textit{if} and \textit{while}.
Keywords \textit{else} and \textit{else if} are changed into \textit{if}.
Keyword \textit{for} is changed into \textit{while}.
Other keywords, such as \textit{break} and \textit{switch}, are removed.

We do not add array or pointer into the noise paths, which avoids adding bugs accidentally caused by pointers or arrays.
Therefore, the return type and the argument type of a function in the FCG are the basic data types.
If in the original program, the definition of a function contains pointer types, such data types will be substituted with basic data types randomly.
For example, "\textit{char * funa}(\textit{int *p})" will be replaced by "\textit{int funa}(\textit{float p})".
The types \textit{int} and \textit{float} are chosen randomly from all the basic types.

\begin{figure*}[thp!]
    \begin{subfigure}
    [The original FCG.
    The weight of path $A\rightarrow C\rightarrow E$ is five.\label{fcg-no-bug}]{%
    \includegraphics[height=0.29\textwidth]{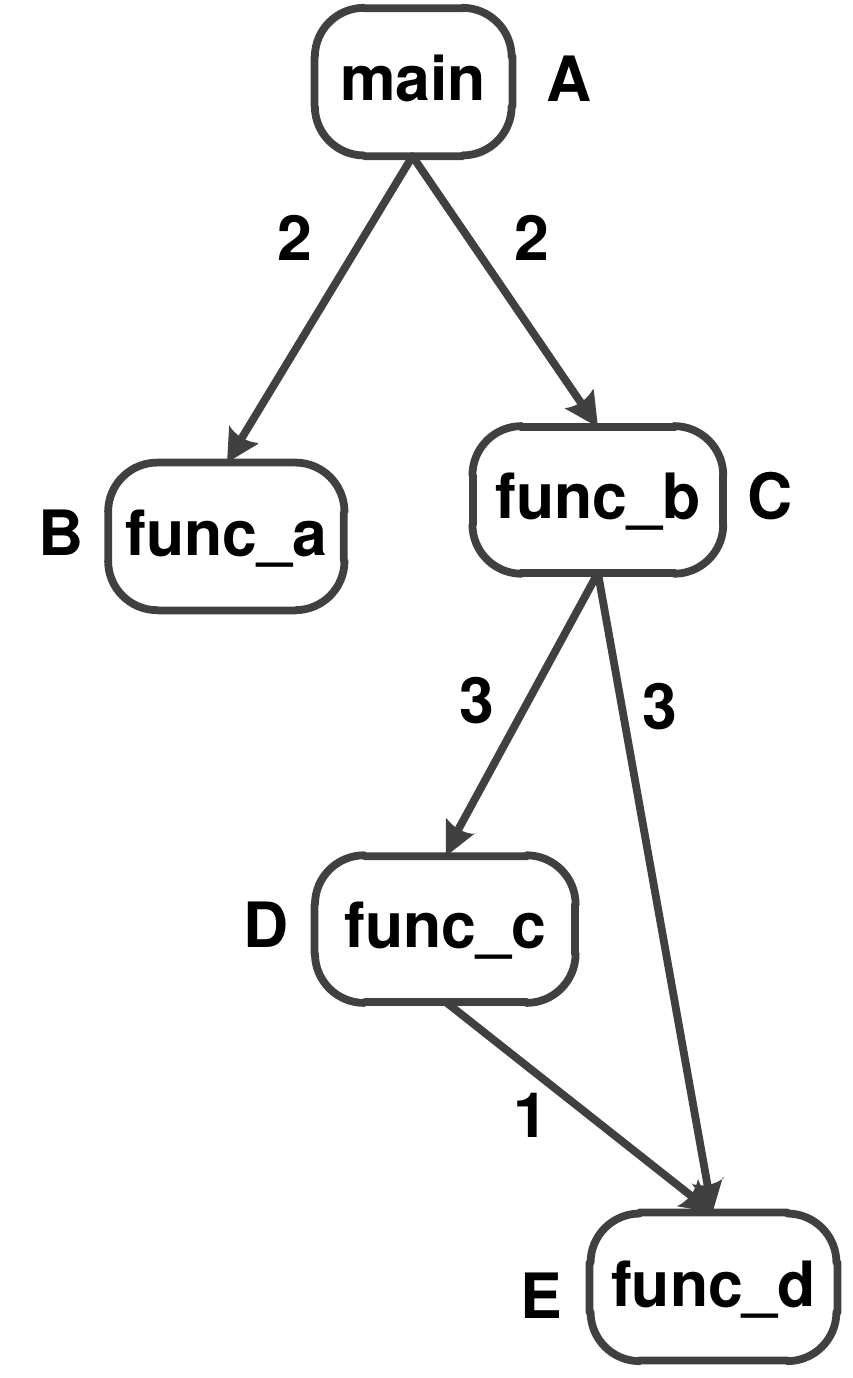}
    } 
    \end{subfigure}
    \hspace{.5in}
    \begin{subfigure}
    [FCG with a bug vertex. 
    The bug node is a potential position to insert a bug.
    The weight of path $A\rightarrow C\rightarrow E\rightarrow F$ is seven.\label{fcg-bug}]{%
    \includegraphics[height=0.29\textwidth]{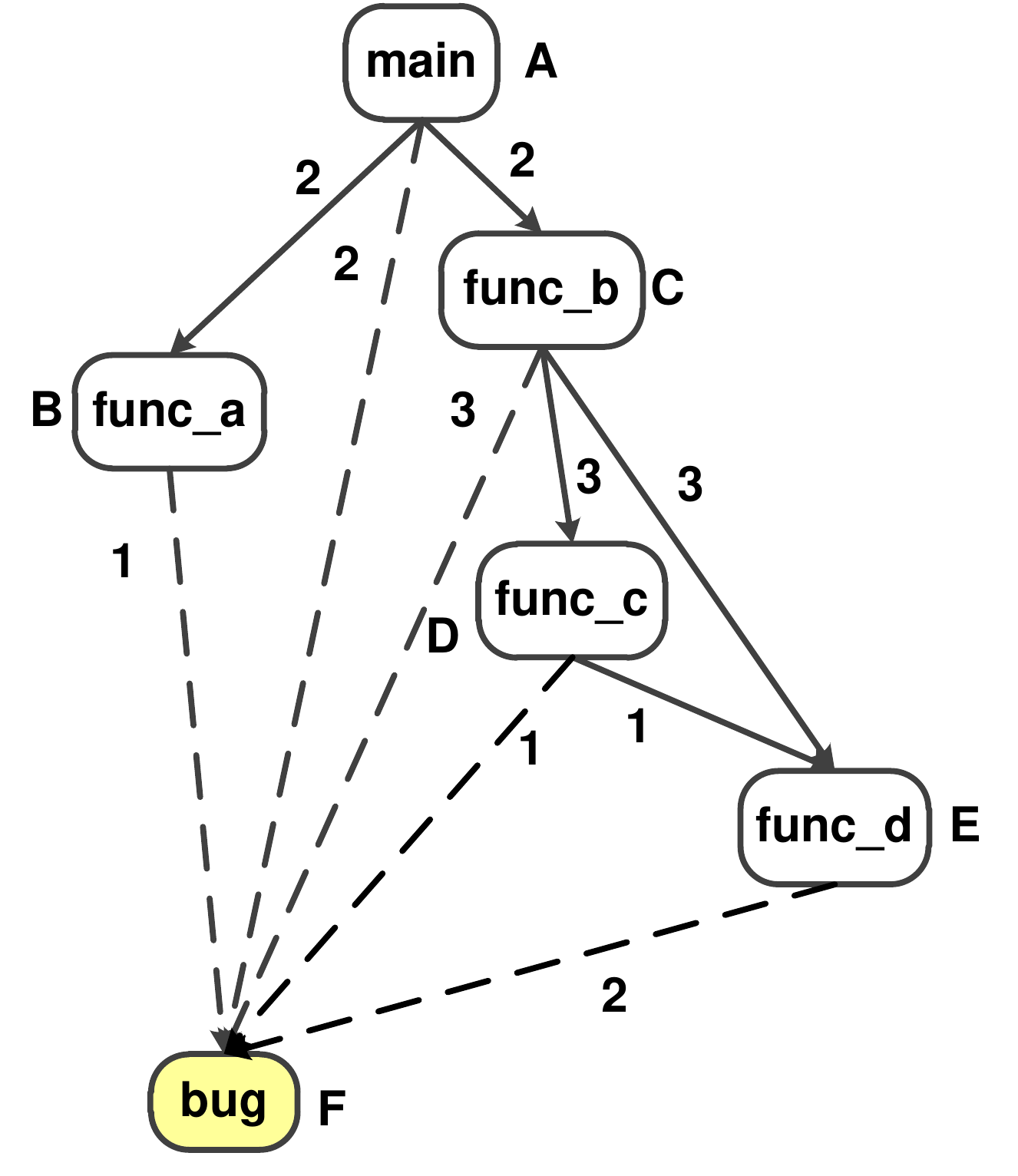}
    }
    \end{subfigure}
    \hspace{.5in}
    \begin{subfigure}
    [The execution paths containing bug path $A\rightarrow C\rightarrow E\rightarrow F$. \label{fcg-bug-dataflow}]{
    \includegraphics[height=0.29\textwidth]{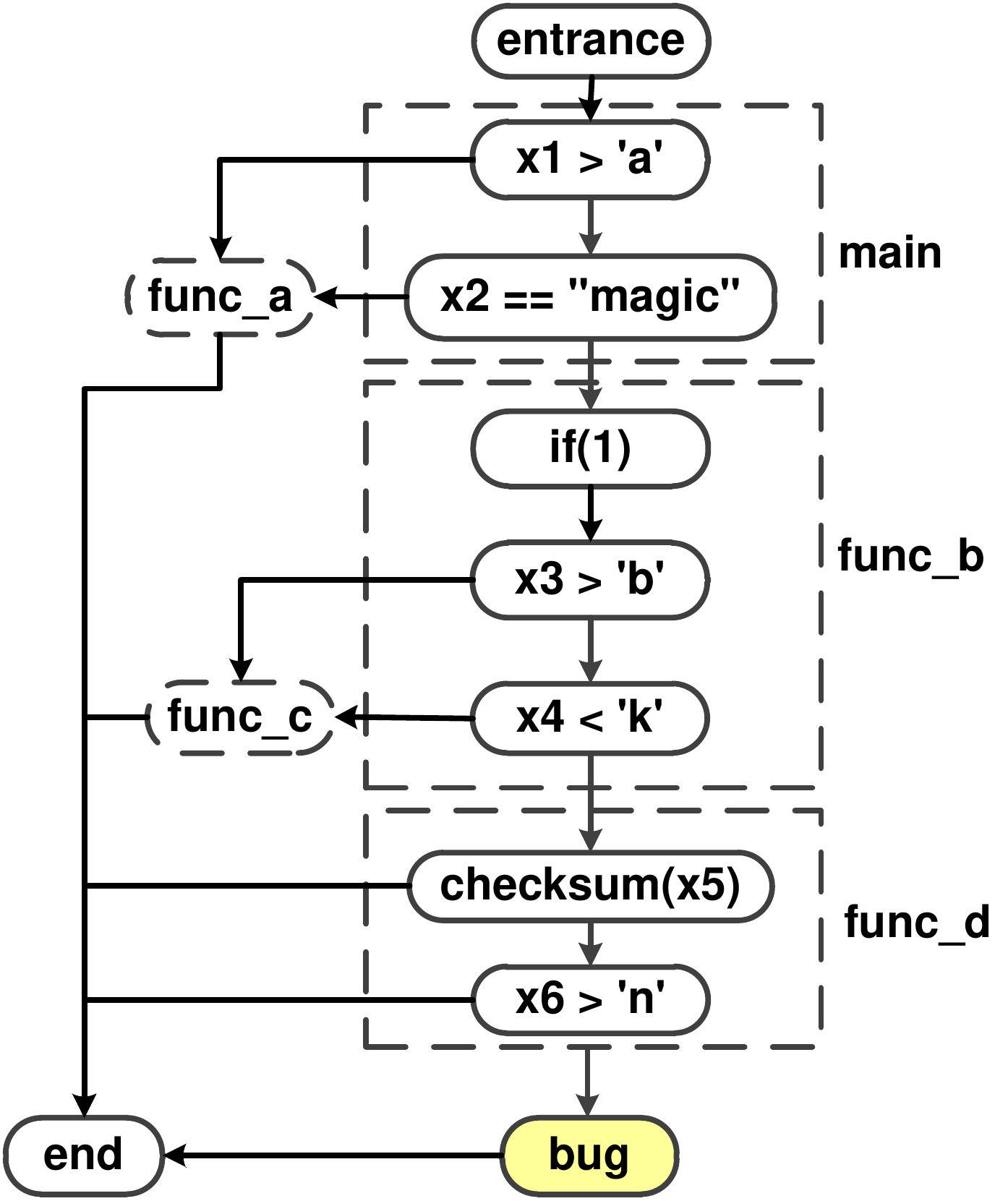} 
    }
    \end{subfigure}
    \caption{The procedure to insert a bug.
    The weight of an edge is the number of \textit{if} statements in the deepest nested condition in the source node.
    Besides, the weight of a path is the sum of all the weights of edges.}
    \label{fcg-bug-no-bug}
\end{figure*}

\subsection{Configure the Number of Features}
Currently, FEData only selects four features as described in Section \ref{sec-Search-hampering}.
After we choose the bug path, different numbers of features can be inserted into it.
Only one dataflow to trigger the inserted bug exists in a program in FEData.
We insert the execution paths that satisfy the four conditions described in Section \ref{sbsec-num-paths}.
All magic values and checksums are inserted into the bug path.
FEData includes three types of programs.
The first type of programs has different numbers of magic values while the number of execution paths is fixed and the number of checksums is zero.
The second includes different numbers of execution paths while the number of magic values and the number of checksums are both zero.
Similarly, the third has programs with different numbers of checksums while the number of execution paths is fixed and the number of magic values is zero.

\begin{lstlisting}[caption={An example of CWE-761.
The key line is the reason for the bug.}, label=cwe761, escapechar=@, float=thp!]
char* vul_var;
vul_var=(char*)malloc(20*sizeof(char));
strcpy(vul_var, "CWE-761");
@\colorbox{light-gray}{vul\_var = vul\_var + 1;}@ //the key line
free(vul_var);
\end{lstlisting}

\subsection{Insert a Unique Bug}
Based on FCG, FEData can finally insert a bug.
FEData first chooses one path in FCG as the bug path.
Then it sets conditions for the bug path.
Finally, at the end of the bug path, it inserts a bug.

\subsubsection{Choose the bug path} \label{choose_bugP}
In the FCG, FEData chooses one path as the bug path.
We first scan all functions and choose the nested conditions with the maximal number of \textit{if} statements in each function.
The weight of an edge in FCG is defined as the number of \textit{if} statements in the chosen nested conditions in the source node.
The weight of a path is the sum of all the weights of edges in the path.
In Fig.\ref{fcg-no-bug}, the function \textit{main} has nested conditions and the maximal number of \textit{if} statements in the nested condition is two, thus the weights of edges \textit{AB} and \textit{AC} are two.

FEData adds a node called bug node into the FCG, and all the nodes in the original FCG are connected to the bug node.
We choose the bug path based on the new FCG.
Recall that in FEData, the number of execution paths equals the number of input-related control statements in the bug path plus one (\textit{i.e. p = c + 1}).
Therefore, if the number of execution paths is set as seven in the bug path, we should choose a path that has more than six \textit{if} statements.
FEData finds all the paths in FCG whose weights are equal to or larger than six and then it picks one of them randomly as the bug path.
In Fig.\ref{fcg-bug}, three paths $A\rightarrow C \rightarrow E\rightarrow F$, $A\rightarrow C\rightarrow D\rightarrow F$ and $A\rightarrow C\rightarrow D\rightarrow E\rightarrow F$ satisfy the requirement, and FEData picks the path $A\rightarrow C\rightarrow E\rightarrow F$ as the bug path randomly.
Finally, we will ensure that only the bug path can reach the bug node, \textit{i.e.} all the noise paths cannot reach the bug node.
This step ensures that only one bug path exists in a program.
Moreover, it also keeps simplicity of counting the number of paths.
For example, the path $A\rightarrow C\rightarrow D\rightarrow E\rightarrow F$ in Fig.\ref{fcg-bug} is set to be infeasible.

\subsubsection{Set Conditions for bug path}

FEData adds conditions in the bug path.
In Section \ref{choose_bugP}, the number of execution paths is set as seven.
Therefore, the number of input-related conditions is six.
If the number of magic values and the number of checksums are both set to one, FEData will randomly pick two of the six \textit{if} statements, one for magic value and another for checksum.
The remaining four conditions are set to normal checks.

In Fig.\ref{fcg-bug-dataflow}, FEData picks the second condition to insert a magic value `magic', and chooses the sixth condition to add the checksum function.
The remaining conditions are larger than or smaller than or always true.
Note that, the weight of edge \textit{CE} is three, but we only need two statements to insert input-related conditions.
Therefore, one of the three conditions is set to \textit{if}(1).
All the variables from \textit{x1} to \textit{x6} get their values from inputs.

\subsubsection{Insert a Bug}\label{ssb-bug-insert}
At the end of the bug path, FEData inserts a bug to the program.
Many kinds of bugs are listed on CWE\footnote{https://cwe.mitre.org}, and FEData chooses the kinds of bugs which can be inserted in key line mode.
These bugs can be fixed by removing or changing one line in the programs.
For example, most bugs caused by overflow are in this category.
Meanwhile, FEData can add different bugs to programs as long as the bugs are in key line mode.
FEData can also be extended to other kinds of bugs, such as logic bugs \cite{Dolan-Gavitt2016LAVA}, because it produces programs from structures of existing programs.

In this paper, FEData inserts a bug, called `Free of Pointer not at Start of Buffer' (\textit{i.e.} CWE-761), from CWE.
This bug occurs when a pointer to a memory resource allocated on the heap is released, but the pointer is not at the start of the buffer.
Listing \ref{cwe761} is a simple example of CWE-761.
The key line is the code line that moves the pointer to the next heap.

\section{Evaluation} \label{sec-eval}
Experiments are run on AMD Opteron 6320, and each fuzzer is run on one core.
We typically run two fuzzers, AFL and AFLFast \cite{Boehme2016Coverage}.
We cannot test other fuzzers because they are unavailable (such as CollAFL \cite{Gan2018CollAFL} and Steelix \cite{Li2017Steelix}) or cannot be appropriately run (such as Driller \cite{Stephens2016Driller}, T-Fuzz \cite{Peng2018T} and Vuzzer \cite{Rawat2017Vuzzer}).
AFL is a greybox fuzzing regarded as the baseline by many other fuzzers.
AFL initially sets some seeds and produces many inputs mutated from these seeds.
Then, those inputs will be fed into the testing program, and new seeds will be selected from the inputs based on the results of the execution.
More inputs will be generated from the new seeds and those inputs will also be fed into the testing program.
This procedure will be processed repeatedly.
A cycle is done when all the seeds are chosen and mutated to produce inputs, and all the generated inputs have been fed into the testing program.
Then, a new cycle begins.
Many fuzzers, including AFLFast, develop their own algorithms based on AFL \cite{Wang2018SAFL,Schumilo2017kAFL,Stephens2016Driller,Boehme2016Coverage,Gan2018CollAFL}.

AFL and AFLFast are evaluated on two experiments.
The first one is designed to assess the ability of the two fuzzers to search specific execution paths.
In this experiment, 90 binaries, which contain different numbers of execution paths and have no specific checks (\textit{i.e.,} magic value or checksum) in the bug path, are used to evaluate AFL and AFLFast.
We stop fuzzing when it has found the inserted bug or it has run for 12 hours.
Another experiment is designed to assess the ability of fuzzers to pass magic values.
In this experiment, 15 different binaries are chosen from FEData to evaluate AFL and AFLFast.
Each of the fifteen binaries has twenty execution paths, one magic value and zero checksums in the bug path.
The length of a magic value is between one and three bytes.
Again, we stop fuzzing when it has found the inserted bug or it has run for 12 hours.
We do not set different numbers of magic values to evaluate AFL and AFLFast because the two fuzzers are not designed to resolve magic values.

As for the policies used for evaluation, we set the same seed, use statistical results, and choose the same programs for both AFL and AFLFast.
Specifically, both of the two fuzzers use Hello World as the original seed.
Besides, we use statistical results to evaluate fuzzers, \textit{i.e.,} the average time of finding a bug.
Moreover, these two fuzzers are evaluated on the same programs and set the same timeout.
We utilize four metrics, including the execution speed, the number of bugs found, the number of identified execution paths\footnote{Note that, AFL can count execution paths but it cannot count them accurately \cite{Gan2018CollAFL}.}, and the number of cycles that have been executed.

\subsection{Experiment I} \label{expm-one}
\begin{figure}[thp!]
\includegraphics[width=1.0\columnwidth]{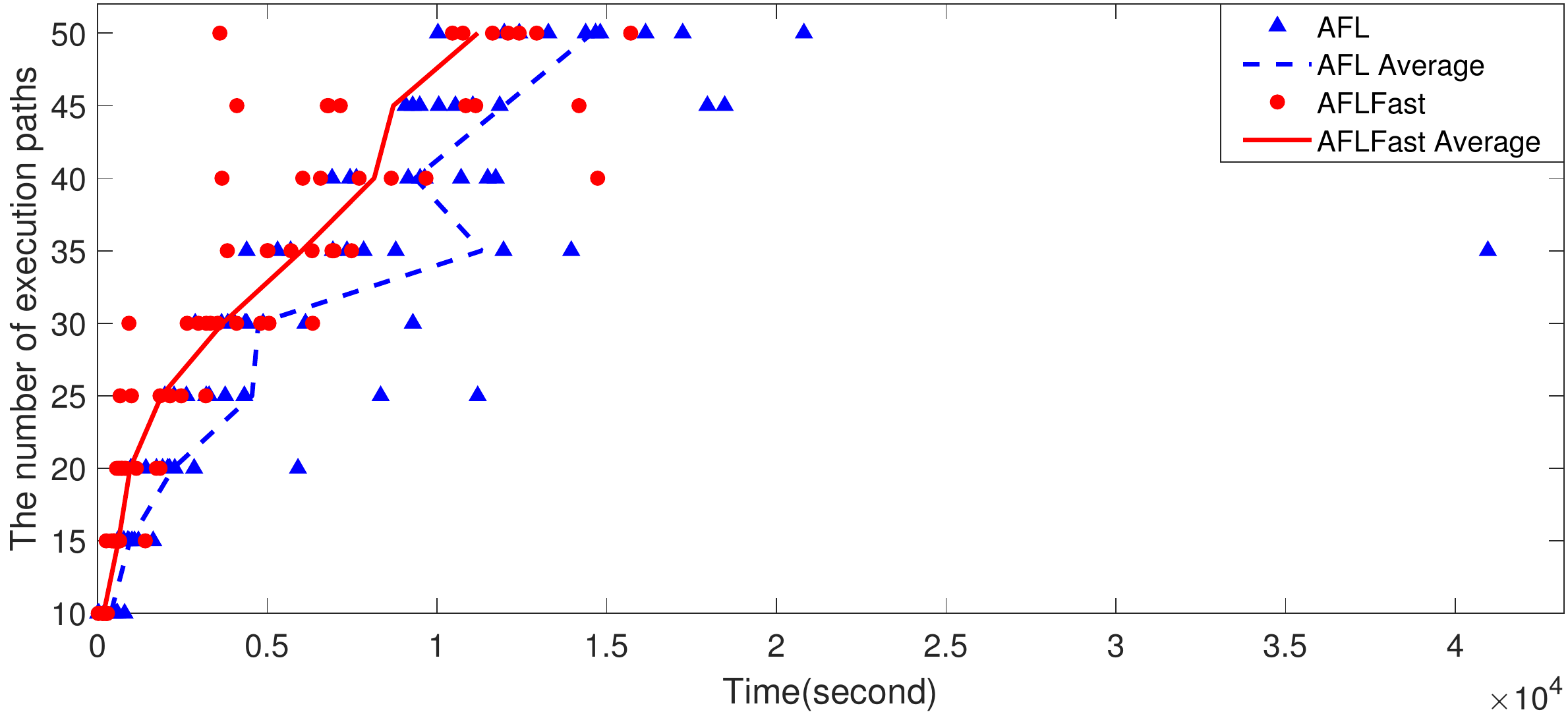}
\caption{AFL and AFLFast are run on 90 binaries.
One dot or triangle indicates an identified bug.}\label{expm-ninty-bins}
\end{figure}

During this experiment, we are the first to find that AFLFast may run too many cycles, more than what it needs.
This is because AFLFast no longer produces inputs and runs through a cycle very fast.
We call this phenomenon \texttt{cycle explosion}. The cycle explosion is a state that a fuzzer can no longer produce new inputs and will run a large number of cycles.
The results are shown in Fig.\ref{expm-ninty-bins} and Table \ref{table-bugs-afl-aflfast}.

\begin{table}
\caption{The Number of Bugs Found in Experiment I}
\label{table-bugs-afl-aflfast}
\begin{minipage}{\columnwidth}
\begin{center}
\begin{tabular}{ccc}
  \toprule
  The number of  & The number of bugs  & The number of bugs \\
    execution paths & found by AFL & found by AFLFast\\
  \midrule
  10 & 10 & 10 \\
  15 & 10 & 10 \\
  20 & 10 & 10 \\
  25 & 9 & 6 \\
  30 & 10 & 10 \\
  35 & 9 & 9 \\
  40 & 9 & 7 \\
  45 & 9 & 7 \\
  50 & 10 & 8 \\
  \bottomrule
\end{tabular}
\end{center}
\bigskip
\footnotesize\emph{Note:} {For each program, ten different binaries are included and each binary has only one bug.}
\end{minipage}
\end{table}

In Fig.\ref{expm-ninty-bins}, either a dot or a triangle denotes a bug found by fuzzing.
The result indicates that AFLFast indeed finds bugs faster than AFL.
Fig.\ref{expm-ninty-bins} also shows that AFLFast searches execution paths faster than AFL in each kind of programs.
When the number of execution paths grows, it costs both AFL and AFLFast more time to trigger bugs.
This trend supports that the number of execution paths affects the time for fuzzing to find bugs.

However, the results in Table \ref{table-bugs-afl-aflfast} tells a different story.
AFLFast finds fewer bugs than AFL, especially when the number of execution paths grows larger.
G. Klees \textit{et al.} also give a similar conclusion that AFLFast may perform worse than AFL when evaluating longer than 24 hours.
We explore the reason for this conclusion and find an interesting phenomenon.
The reasons that AFLFast cannot find some bugs are different from AFL.
All the four bugs that AFL cannot find is due to the limited time.
However, all the 13 bugs that AFLFast cannot find are because of cycle explosion.
Based on Fig.\ref{expm-ninty-bins} and Table \ref{table-bugs-afl-aflfast}, the conclusion is that AFLFast finds bugs faster than AFL, but it has a chance to get trapped in the cycle explosion, especially when the bugs are protected in a deep path.

\subsection{Experiment II}
\begin{figure}[t]
    \begin{subfigure}
    [The execution speed (per second: /s). \label{magic-fif-execs}]{
    \includegraphics[width=0.95\columnwidth]{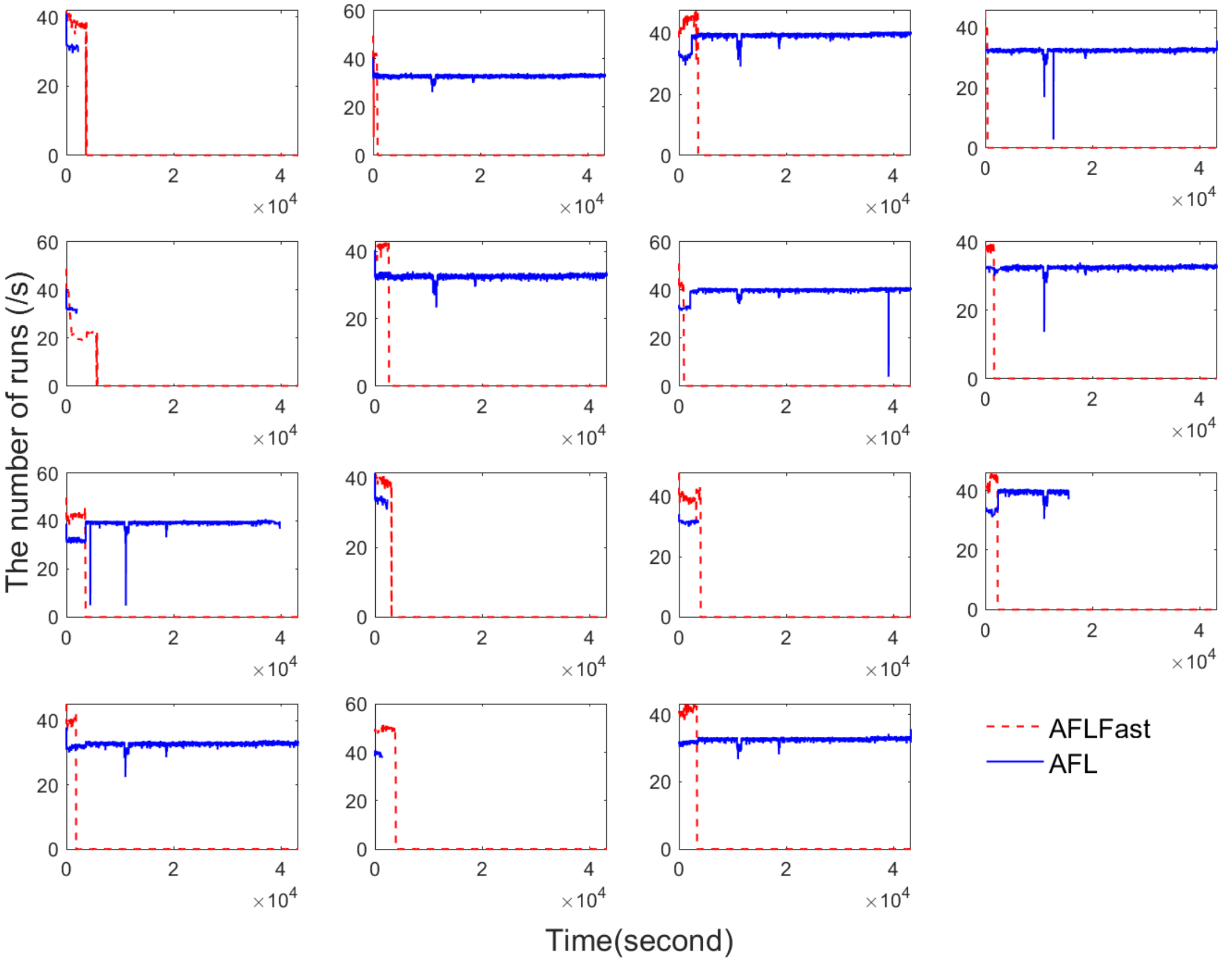} 
    }
    \end{subfigure}
    \begin{subfigure}
    [The number of cycles that finished.
    \label{magic-fif-cycles}]{
    \includegraphics[width=0.95\columnwidth]{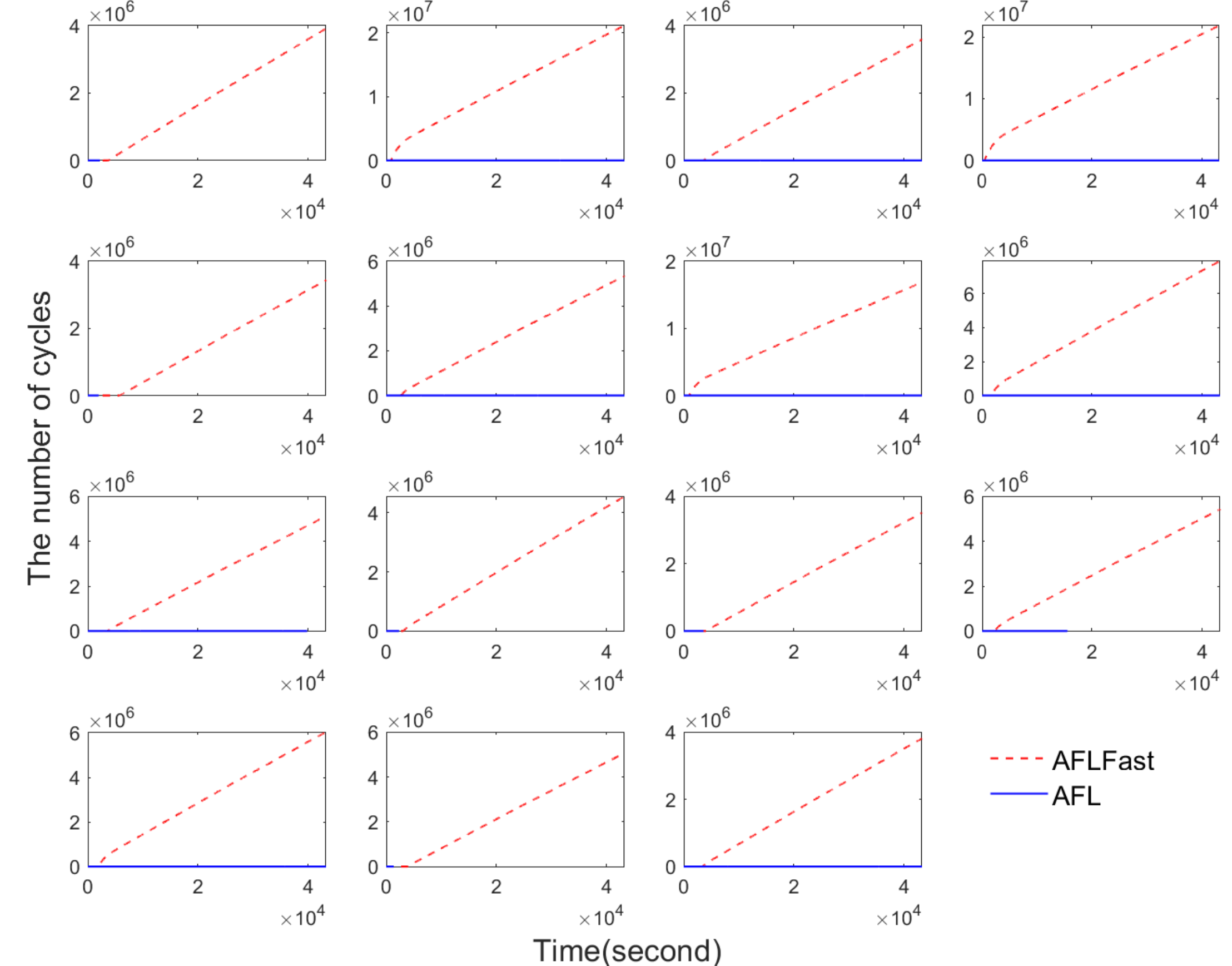}
    }
    \end{subfigure}
    \caption{Evaluation results of AFLFast and AFL on 15 binaries.
    Each binary contains one magic value in the bug path.}\label{magic-fifs}
\end{figure}

In this experiment, AFL found seven bugs while AFLFast found none.
Fig.\ref{magic-fifs}(A) shows the execution speed and Fig.\ref{magic-fifs}(B) shows the number of cycles that have been run.
As shown in Fig.\ref{magic-fifs}, AFLFast gets trapped in the cycle explosion quickly.
As for AFLFast, the execution speed drops to 0/s quickly while the number of cycles rises to millions.
Before the 10,000th second, AFLFast has already been in cycle explosion on all the 15 binaries.
After that, AFLFast can no longer execute binaries and will run through a cycle very quickly.
Therefore, it cannot find bugs since then.
However, AFL executes binaries in a speed between 20/s and 40/s.
Besides, the number of cycles run by AFL is less than 2000.

When cycle explosion occurs, the number of execution paths found by AFLFast is fewer than 20, which is the number we set for execution paths.
This means AFLFast cannot pass the magic values and stop finding new execution paths.
This experiment shows that AFLFast performs worse than AFL regarding the number of bugs found.
Therefore, AFLFast does not perform better than AFL regarding the magic values.

Based on these two experiments, we conclude that AFLFast does find bugs faster than AFL. However, AFLFast finds fewer bugs than AFL if bugs are time-consuming for fuzzing to find them.

\section{Improvement of AFLFast (AFLFast+)} \label{sec-improve-fast}
It is impressive that the cycle explosion prevents AFLFast from triggering bugs in both experiments.
Therefore, we dig deeper into the reason for it and present a solution to improve AFLFast.

\subsection{An Interesting Case}
First of all, one interesting case from Section \ref{expm-one} is analyzed.
We choose a program called "terminology\_0\_50\_", which has 50 execution paths in the bug path.
There are no magic values and checksums in the bug path.
AFL finds the bug at the 13380th second.
AFLFast cannot find the bug even at the 31380th second.
Note that we run AFLFast on its default mode, \textit{i.e.}, the FAST mode.

\begin{figure*}[!htb]
\includegraphics[width=0.98\textwidth]{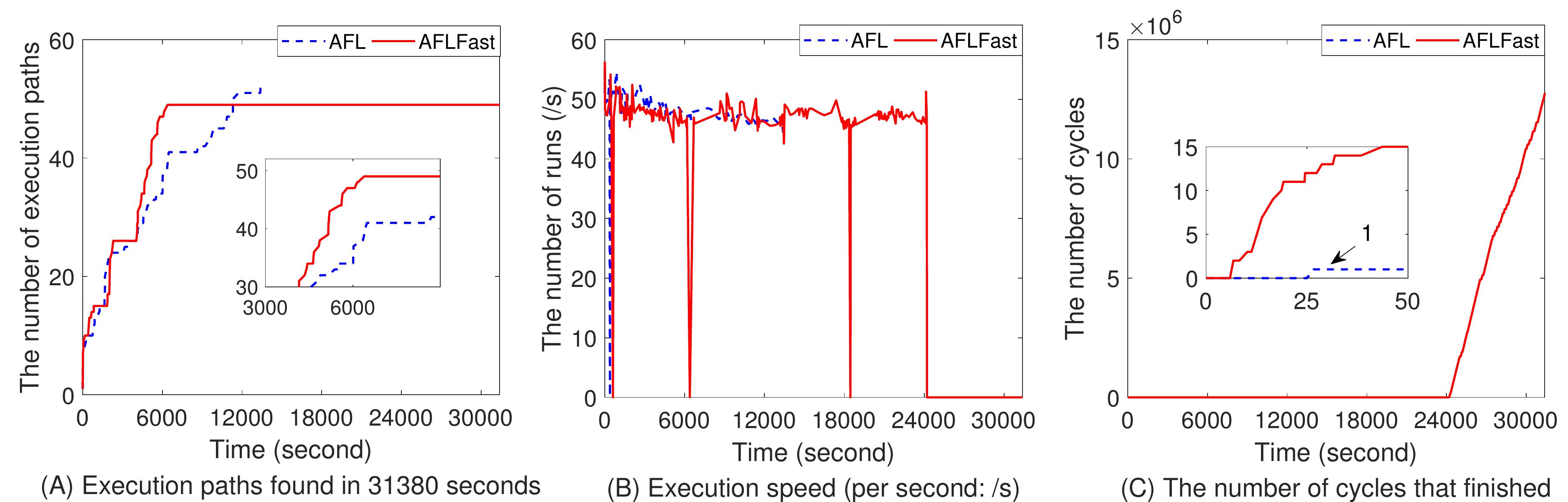} 
\caption{An interesting case.
    In the first 3600 seconds, AFL and AFLFast find execution paths at almost the same speed.
    From the 3600th second to the 6360th second, AFLFast finds execution paths faster than AFL.
    The interesting part is that AFLFast cannot find more execution paths since the 6360th second while AFL finds more execution paths until it crashes the program.}
    \label{afl-vs-aflfast}
\end{figure*}

The results in Fig.\ref{afl-vs-aflfast} show the strength and weakness of AFLFast.
In the first 3600 seconds, as shown in Fig.\ref{afl-vs-aflfast}(A), the speed of finding execution paths are similar.
The performance of AFL and AFLFast are close at first.
This is because AFLFast has to build its Markov Chain Model at first, in which the transition possibilities are set during runs.
When most transition possibilities of the Markov Chain Model are set, AFLFast starts to perform better than AFL, which is the strength of AFLFast.

The interesting part starts at the 6360th second.
Since the 6360th second, AFLFast stops finding more execution paths, but AFL still searches for more execution paths.
AFL and AFLFast intersect at the 11280th second, and AFL finally crashes the program at the 13380th second.
The nature of AFLFast is to produce inputs that prefer less-frequent execution paths, but less-frequent execution paths may bring fuzzing far from the bug.
In this binary, the bug path is extremely deep, and it will be hit the most often if the bug position is reached.
This explains that AFLFast cannot reach more execution paths.
Meanwhile, at the 6360th second and the 18420th second, the execution speed is very low, which is shown in Fig.\ref{afl-vs-aflfast}(B).
Especially after the 24180th second, the execution speed stays at 0/s.
According to AFLFast, the number of inputs generated for examining the binary is inversely proportional to the number of times that one execution path is exercised.
Therefore, when most execution paths in the program are exercised for a large number of times, the execution speed will be very low.
To make the situation worse, because each seed is assigned to a very low number of inputs, fuzzing will go through a cycle very quickly.
The consequence is, the number of cycles done by fuzzing grows to a large number.
This is the cycle explosion.
In Fig.\ref{afl-vs-aflfast}(C), AFLFast runs more than 10 million cycles after the 24180th second, which comes along with 0 input being produced.
It explains the reason why the execution speed is 0/s after the 24180th second in Fig.\ref{afl-vs-aflfast}(B).
On the other hand, AFL almost stays executing binaries at the speed about 50/s.

\subsection{AFLFast+}
AFL assigns almost constant energy (\textit{i.e.,} the number of inputs generated from seeds) to each seed.
Therefore, AFL does not cause cycle explosion and even can find bugs in one cycle.
AFLFast improves AFL by assigning different number of inputs to a seed.
Different mutation strategies, including FAST mode, LINEAR mode, and QUAD mode, are utilized by AFLFast.

The FAST mode assigns energy to state \textsl{i} as
\begin{equation}\label{aflfast-fast}
  p_f(i) = \textrm{min} \bigg(\frac{\alpha(i)}{\beta}\cdot\frac{2^{s(i)}}{f(i)}, M \bigg)
\end{equation}
Where $\alpha(i)$ is the number of inputs generated by AFL, $s(i)$ is the number of times that seed $d_i$ is chosen from the seed queue, $f(i)$ is the number of generated inputs that have exercised state $i$. The constant $M$ provides an upper bound on the number of inputs that are generated per fuzzing iteration.

The LINEAR mode increases the energy of state \textsl{i} linearly with respect to $s(i)$, which is
\begin{equation}\label{aflfast-linear}
  p_l(i) = \textrm{min} \bigg(\frac{\alpha(i)}{\beta}\cdot\frac{s(i)}{f(i)}, M \bigg)
\end{equation}

The QUAD mode is based on the LINEAR mode, and the energy of fuzzing is computed as
\begin{equation}\label{aflfast-quad}
  p_d(i) = \textrm{min} \bigg(\frac{\alpha(i)}{\beta}\cdot\frac{s(i)^{2}}{f(i)}, M \bigg)
\end{equation}

All these three power schedules assign energy to state $i$ inversely proportional to $f(i)$, which is the reason for cycle explosion.
When $f(i)$ grows to a large value, the energy is very low.
To improve AFLFast, one way is to set a lower bound to its energy strategies.
\textsl{L} is a lower bound set for AFLFast.
Therefore, the FAST mode is improved as
\begin{equation}\label{aflfast-fast-imp}
  q_f(i) = \textrm{max}\big(p_f(i), L \big)
\end{equation}

For the LINEAR mode, we have
\begin{equation}\label{aflfast-linear-imp}
  q_l(i) = \textrm{max}\big(p_l(i), L \big)
\end{equation}

And for the QUAD mode, we also have
\begin{equation}\label{aflfast-quad-imp}
  q_d(i) = \textrm{max}\big(p_d(i), L \big)
\end{equation}

\subsubsection{Evaluation about Cycle Explosion}
\begin{figure*} 
\includegraphics[height=0.28\textwidth]{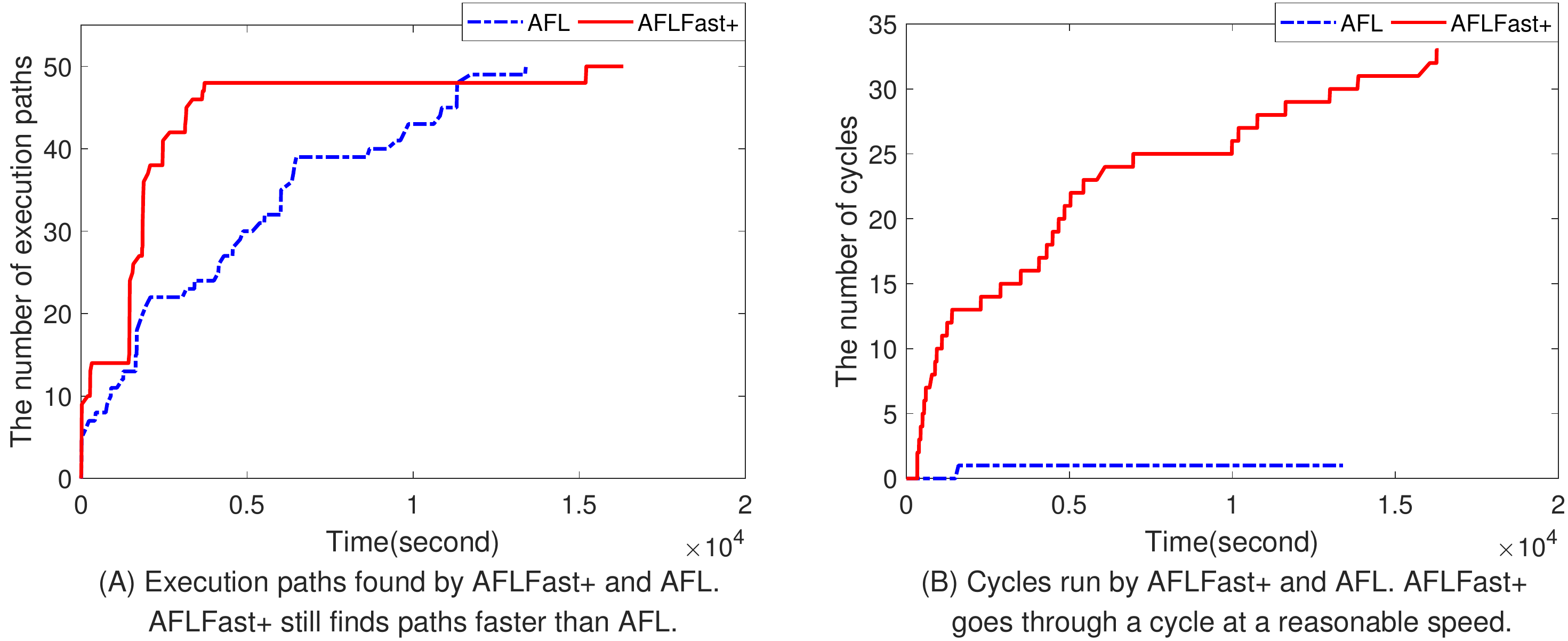}
\caption{Run AFLFast+ on "terminology\_0\_50\_".}
\label{afl-new-aflfast}
\end{figure*}

With the AFLFast+, the results of testing the interesting case are shown in Fig.\ref{afl-new-aflfast}.
The bug, which is not found by AFLFast, is exposed by AFLFast+. Meanwhile, AFLFast+ still searches execution paths faster than AFL.
The number of cycles run by AFLFast+ is in a reasonable range.

\begin{figure*}[!htb]
\includegraphics[height=0.28\textwidth]{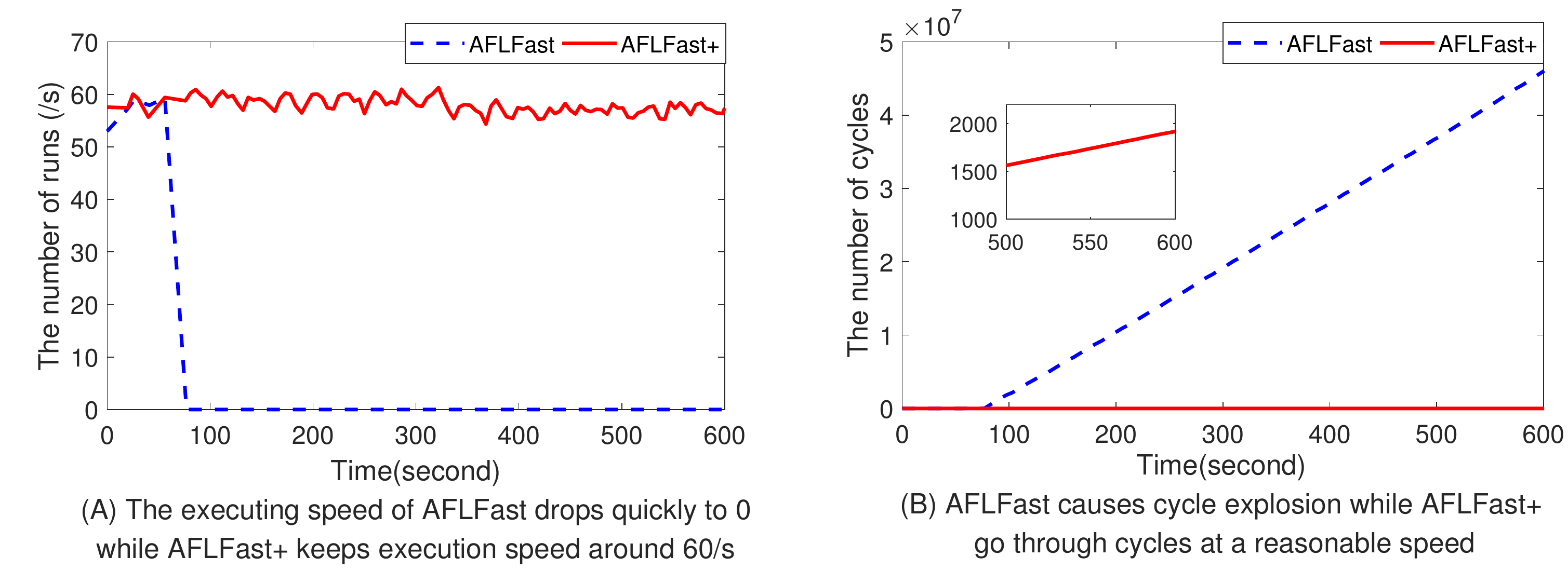}
\caption{Cycle Explosion. AFLFast and AFLFast+ are run on "BMFS\_31\_1\_10".}
\label{aflfast-newfast}
\end{figure*}

To research on the issue about cycle explosion of AFLFast and AFLFast+, a different program, "BMFS\_31\_1\_10", is chosen.
The new program only has ten execution paths but contains a magic value in the bug path.
We run AFLFast and AFLFast+ on this program for ten minutes.
During the experiment, only one execution path is found by both fuzzers because the magic value prevents them from searching for more execution paths.
This is the worst case for AFLFast, in which all the genetated inputs exercise the same execution path.
Therefore, there are few or no new inputs will be generated by AFLFast.
However, as for AFLFast+, it has a lower bound to avoid producing no inputs.
The number of cycles and the execution speed are shown in Fig.\ref{aflfast-newfast}.
AFLFast quickly falls into the cycle explosion while AFLFast+ runs a cycle at a reasonable speed.
In Fig.\ref{aflfast-newfast}, the execution speed of AFLFast decreases fast and stays at 0/s after the 77th second.
However, the execution speed of AFLFast+ is around 60/s throughout ten minutes.

\subsubsection{Evaluation On the 105 Binaries.}
\begin{figure}[!htb]
\includegraphics[width=1.0\columnwidth]{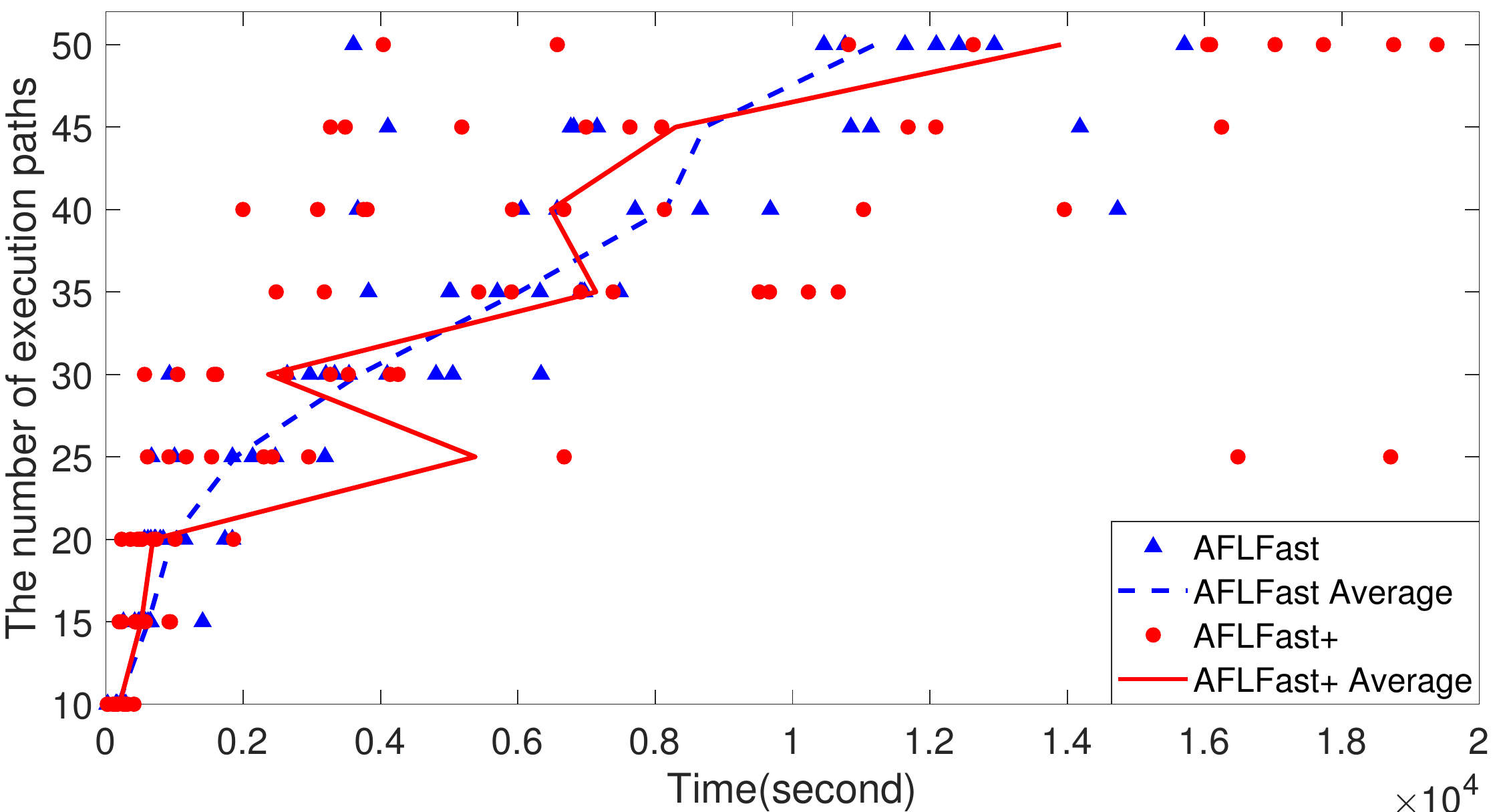}
\caption{The results of AFLFast and AFLFast+ which are tested on 90 binaries.
One dot or triangle is a bug found.}\label{expm-plus-ninty}
\end{figure}

We evaluate our AFLFast+ on the 90 binaries in Experiment I.
Fig.\ref{expm-plus-ninty} shows that the efficiency of AFLFast and AFLFast+ is almost the same.
However, AFLFast+ finds 88 bugs while AFLFast finds only 77 bugs, which is shown in Table \ref{table-bugs-fast-fastp}.
Meanwhile, the reason of the two bugs not being found by AFLFast+ is due to the limited time, and the cycle explosion does not appear.
We also run AFLFast+ on the 15 binaries in Experiment II.
Five bugs are detected by AFLFast+. Recall that, in Experiment II, AFLFast finds no bugs.
Therefore, AFLFast+ keeps the efficiency but can detect more bugs than AFLFast.

\begin{table}
\caption{Bugs Found by AFLFast+ and AFLFast}
\label{table-bugs-fast-fastp}
\begin{minipage}{\columnwidth}
\begin{center}
\begin{tabular}{ccc}
  \toprule
  The number of  & The number of bugs  & The number of bugs \\
    execution paths & found by AFLFast+ & found by AFLFast\\
  \midrule
  10 & 10 & 10 \\
  15 & 10 & 10 \\
  20 & 10 & 10 \\
  25 & 10 & 6 \\
  30 & 10 & 10 \\
  35 & 10 & 9 \\
  40 & 9 & 7 \\
  45 & 9 & 7 \\
  50 & 10 & 8 \\
  \bottomrule
\end{tabular}
\end{center}
\bigskip
\footnotesize\emph{Note:} {For each program, ten different binaries are included and each binary has only one bug.}
\end{minipage}
\end{table}

\section{Related Work} \label{sec-related-work}

\subsection{Corpora}
There is not much work about synthesizing corpora for fuzzing.
The closest research is LAVA.
The main idea of LAVA is to insert a bug into a real-world program, and the synthetic program keeps the contexts of real-world programs.
LAVA uses taint analysis to search DUAs (Dead, Uncomplicated and Available data) in programs, which is the key to maintain the realness.
Based on the DUAs, LAVA inserts some bugs and adds magic values to protect the bugs.
As it is based on taint analysis, the number of potential bugs to be inserted depends on the knowledge about the target real-world programs, which limits LAVA to insert more bugs.
Meanwhile, LAVA offers dataflow to trigger bugs, but the details about the bug path are ambiguous.
LAVA synthesizes a program based on the idea that the inserted bugs can be found in real-world programs.
Although LAVA-M argues the realness of its generated programs, recent experiment\footnote{http://moyix.blogspot.com/2018/03/of-bugs-and-baselines.html} shows that techniques following a simple pattern effectively find all of the LAVA-M bugs.
Another synthetic corpus is DARPA CGC, which includes hundreds of programs.
DARPA's competitors test their algorithms on CGC and attempt to discover as many bugs as possible in 24 hours.
CGC has the same weakness as LAVA that features preventing fuzzers from finding bugs are not quantified.
Moreover, the NIST Software Assurance Metrics And Tool Evaluation (SAMATE) project assembles a large public corpus, including 86,864 synthetic C and Java programs that exhibit 118 different CWEs.
However, programs in this corpus contains few search-hampering features.
Therefore, it is impossible to evaluate fuzzing individually on these features.

Most fuzzers \cite{Peng2018T,Chen2018Angora,Gan2018CollAFL,Chen2018Hawkeye,Boehme2017Directed,Schumilo2017kAFL,Wang2017Skyfire,Rawat2017Vuzzer,Stephens2016Driller} evaluate their performance on real-world programs that is manually collected.
Such programs usually do not provide with information about bugs, and some even do not know whether there exist bugs.
However, almost none of them use the same programs if their versions are considered.
This makes it difficult to compare the performance of fuzzers.
The collection of real-world programs is tedious and painstaking.
The Google Fuzzer Test Suite is a collection of real-world programs.
To the day the paper is written, 25 programs are collected.
This corpus has information about bugs but does not include the details about features.

\subsection{Fuzzing Algorithms}

Many fuzzers are designed to present solutions for the three challenges, including checksum, magic value, and execution path search.
Coverage-guided fuzzers try to reach as much code coverage as possible.
Some papers try to reach larger coverage by requiring inputs to comply with a specific data format (such as size, type, and checksum).
MoWF \cite{Pham2016Model} and Skyfire \cite{Wang2017Skyfire} extract specific data format from existing corpora.
Difuze \cite{Corina2017Difuze} uses static analysis to compose correctly-structured inputs.
TaintScope \cite{Wang2010TaintScope} leverages a checksum detection algorithm to remove checksums.
Fuzzers try their efforts on satisfying the data format that programs require.

When a fuzzer confronts magic values, it has to satisfy all the bytes otherwise it fails the check and cannot run the code protected by them.
Therefore, if a bug is protected by magic values, a fuzzer has to figure out the specific values, and then the bug can be triggered.
Many papers propose different algorithms to resolve magic values.
Steelix \cite{Li2017Steelix} instruments the binary to provide coverage and comparison progress information and infer the location information of magic values.
Such information is then used to guide the mutation.
T-Fuzz \cite{Peng2018T} flips the conditions for conditional jumps in order to pass the specific checks.
That is, when fuzzing gets stuck, T-Fuzz negates the condition to help pass the check.
Driller \cite{Stephens2016Driller} uses symbolic execution to resolve the magic value.
Angora \cite{Chen2018Angora} mutates only the bytes that solve the path constraints instead of the entire inputs.

The number of execution paths in the program affects the efficiency of fuzzing.
If a program only has one execution path, a fuzzer can go through the program and find the bug quickly (no checks about magic values and data format in the program).
AFLFast \cite{Boehme2016Coverage} takes Markov chain as the model of transitions between states and leverages this model to guide fuzzing to generate inputs that prefer less-frequent execution paths.
CollAFL \cite{Gan2018CollAFL} improves AFL via mitigating path collisions. It changes the computation of hash for edges, which improves the ability to differentiate paths.
Vuzzer \cite{Rawat2017Vuzzer} uses static analysis to model the control flow graph as Markov model, based on which it tries to guide fuzzing to generate inputs to reach larger code coverage.

\section{Discussion: Threat to Validity} \label{sec-discussion}

Although FEData works well on AFL and AFLFast, it can be further improved.
Issues such as the ways to evaluate FEData and improve FEData can be further discussed.

\textbf{The Variety of Fuzzers.}
We only run AFL and AFLFast on FEData because other fuzzers are unavailable or cannot be appropriately run.
In order to further assess FEData, different fuzzers should be run on FEData.
Different fuzzers focus on solving different challenges.
AFLFast matches the feature of execution path in FEData because AFLFast aims to find execution paths faster than other fuzzers.
Therefore, in our experiments, we run AFLFast on 90 binaries which have a different number of execution paths.
Besides, the feature of magic value is also validated by AFLFast, to some extent, because it indeed performs worse than AFL on some programs of FEData.
In order to assess the feature of magic value in FEData, a better way is to use fuzzers trying to help resolve magic values, such as Driller and T-Fuzz.
Similarly, it is better to assess the feature of checksum in FEData by fuzzers focusing on checksums.
We will continue our investigation to run more fuzzers on FEData.

\textbf{The Realness of FEData.}
To customize search-hampering features in the contexts of bugs, FEData sacrifices some realness of the generated programs.
To make FEData more similar to real-world programs, one way is to add functionality by functional programming.
For example, it does not affect a program if the functions in the noise paths change.
Therefore, we can add functionality into such functions.
Another way to improve the realness of FEData is to research the contexts behind bugs.
And such contexts can be added to FEData when a bug is inserted.

\textbf{Issues About the Inserted Bugs.}
Bugs in FEData are inserted in a specific pattern.
Therefore, similarly to LAVA, some simple algorithms may find all the inserted bugs.
In fact, more efforts need to be made on this issue, such as how to insert bugs more complexly.
The directed fuzzers first search some targets, among which they hope bugs are included.
To make it complicated for searching targets, FEData can add bug-free code that is similar to the inserted bug, such as adding code without the key line of bugs.
Another problem is about the unique bug, which is only one in one program.
This may cause problems because the code coverage detected by fuzzers is not precise.
Bugs may be invisible if different coverage is identified as the same one.
Therefore, the bug in a program may not be triggered at all.
To fix this, fuzzers have to develop new algorithms for more precise coverage.
As to the usage of FEData, the result will be more convincing if fuzzers are evaluated on a number of binaries.

\section{Conclusion} \label{sec-conclusion}
In order to evaluate fuzzing more specifically and effectively, we propose generating corpora based on search-hampering features.
Further, we design a prototype corpus, FEData, to show the effectiveness of our idea.
While other corpora focus on bugs, FEData focuses on the search-hampering features.
FEData tests the ability of fuzzers to find solutions for challenges, \textit{i.e.,} the dataflow to trigger bugs, execution path search, magic values, and checksums.
Fuzzing is designed to detect bugs, so the dataflow to trigger bugs is a basic feature and can help count the number of bugs accurately.
We extract the other three features from the basic structures of programs.
FEData includes programs that contain a different number of these three features.
Details about the features are described in this paper.

AFL and AFLFast are evaluated on FEData.
The advancement of AFLFast, which is AFLFast finds execution paths faster than AFL, is supported by FEData.
However, the drawback of AFLFast is magnified by some programs in FEData.
We find that AFLFast is prone to be trapped in cycle explosion (for the first time) if bugs are embedded deeply or fuzzers hit some specific execution paths in a large quantity.
Therefore, AFLFast is improved to AFLFast+ via setting a lower bound to its energy strategies.
The results show that AFLFast+ can find more bugs than AFLFast while the efficiency stays the same.


\bibliographystyle{ACM-Reference-Format}
\bibliography{dataset_generation}

\end{document}